\documentclass[preprint2]{aastex}

\usepackage{calrsfs,euscript,mathrsfs,amssymb}
\usepackage{graphicx}
\usepackage[nomarkers,nolists]{endfloat}
\usepackage{lineno}
\usepackage{ulem}
\newcount\Comments  % 0 suppresses notes to selves in text
\usepackage[usenames,dvipsnames]{xcolor}
\definecolor{darkgreen}{rgb}{0,0.5,0}
\definecolor{purple}{rgb}{1,0,1}
\definecolor{darkpurple}{rgb}{0.5,0,0.5}
\definecolor{lightgreen}{RGB}{135,220,0}
\definecolor{darkgreen}{RGB}{0,102,0}

\newcommand{\kibitz}[2]{\ifnum\Comments=1\textcolor{#1}{#2}\fi}
\newcommand{\lat}{\textit{Fermi}-LAT }

\newcommand{\eg}{e.g.,~}
\newcommand{\ie}{i.e.,~}

\newenvironment{changemargin}[2]{%
\begin{list}{}{%
\setlength{\topsep}{0pt}%
\setlength{\leftmargin}{#1}%
\setlength{\rightmargin}{#2}%
\setlength{\listparindent}{\parindent}%
\setlength{\itemindent}{\parindent}%
\setlength{\parsep}{\parskip}%
}%
\item[]}{\end{list}}

\shorttitle{3FHL Catalog}
\shortauthors{Dom{\'i}nguez, Ajello, Benoit, Cutini, Fortin}
\slugcomment{}

\begin{document}


\title{3FHL: The Third Catalog of Hard \lat Sources}
%\email{Version 6.6 ({\it Preliminary})\\June 14, 2017}
%\author{\lat collaboration}
\author{\small{
M.~Ajello\altaffilmark{1,2}, 
W.~B.~Atwood\altaffilmark{3}, 
L.~Baldini\altaffilmark{4}, 
J.~Ballet\altaffilmark{5}, 
G.~Barbiellini\altaffilmark{6,7}, 
D.~Bastieri\altaffilmark{8,9}, 
R.~Bellazzini\altaffilmark{10}, 
E.~Bissaldi\altaffilmark{11,12}, 
R.~D.~Blandford\altaffilmark{13}, 
E.~D.~Bloom\altaffilmark{13}
% ...
}}
\altaffiltext{1}{Department of Physics and Astronomy, Clemson University, Kinard Lab of Physics, Clemson, SC 29634-0978, USA}
\altaffiltext{2}{email: majello@slac.stanford.edu}
\altaffiltext{3}{Santa Cruz Institute for Particle Physics, Department of Physics and Department of Astronomy and Astrophysics, University of California at Santa Cruz, Santa Cruz, CA 95064, USA}
\altaffiltext{4}{Universit\`a di Pisa and Istituto Nazionale di Fisica Nucleare, Sezione di Pisa I-56127 Pisa, Italy}
\altaffiltext{5}{Laboratoire AIM, CEA-IRFU/CNRS/Universit\'e Paris Diderot, Service d'Astrophysique, CEA Saclay, F-91191 Gif sur Yvette, France}
\altaffiltext{6}{Istituto Nazionale di Fisica Nucleare, Sezione di Trieste, I-34127 Trieste, Italy}
\altaffiltext{7}{Dipartimento di Fisica, Universit\`a di Trieste, I-34127 Trieste, Italy}
\altaffiltext{8}{Istituto Nazionale di Fisica Nucleare, Sezione di Padova, I-35131 Padova, Italy}
\altaffiltext{9}{Dipartimento di Fisica e Astronomia ``G. Galilei'', Universit\`a di Padova, I-35131 Padova, Italy}
\altaffiltext{10}{Istituto Nazionale di Fisica Nucleare, Sezione di Pisa, I-56127 Pisa, Italy}
\altaffiltext{11}{Dipartimento di Fisica ``M. Merlin" dell'Universit\`a e del Politecnico di Bari, I-70126 Bari, Italy}
\altaffiltext{12}{Istituto Nazionale di Fisica Nucleare, Sezione di Bari, I-70126 Bari, Italy}
\altaffiltext{13}{W. W. Hansen Experimental Physics Laboratory, Kavli Institute for Particle Astrophysics and Cosmology, Department of Physics and SLAC National Accelerator Laboratory, Stanford University, Stanford, CA 94305, USA}
% ...

\clearpage
\newpage
\pagebreak

\begin{abstract}
We present a catalog of sources detected above 10~GeV by the {\it Fermi} Large Area Telescope (LAT) in the first 7 years of data using the Pass 8 event-level analysis. This is the Third Catalog of Hard {\it Fermi}-LAT Sources (3FHL), containing 1556 objects characterized in the 10~GeV--2~TeV energy range. The sensitivity and angular resolution are improved by factors of 3 and 2 relative to the previous LAT catalog at the same energies (1FHL). The vast majority of detected sources (79\%) are associated with extragalactic counterparts at other wavelengths, including 16 sources located at very high redshift ($z>2$). Eight percent of the sources have Galactic counterparts and 13\% are unassociated (or associated with a source of unknown nature). The high-latitude sky and the Galactic plane are observed with a flux sensitivity of 4.4 to 9.5~$\times 10^{-11}$~ph~cm$^{-2}$~s$^{-1}$, respectively (this is approximately 0.5\,\% and 1\,\% of the Crab Nebula flux above 10~GeV). The catalog includes 214 new $\gamma$-ray sources. The substantial increase in the number of photons (more than 4 times relative to 1FHL and 10 times to 2FHL) also allows us to measure significant spectral curvature for 32 sources and find flux variability for 163 of them. Furthermore, we estimate that for the same flux limit of $10^{-12}$~erg~cm$^{-2}$~s$^{-1}$, the energy range above 10~GeV has twice as many sources as above 50~GeV, highlighting the importance, for future Cherenkov telescopes, of lowering the energy threshold as much as possible.

\end{abstract}


%%%%%%%%%%%%%%%%%%%%%%%%%%%%%%%%%%%%%%%%%%%%%%%%%%%%%%%%%%%%%%%%
%
%         Introduction 
%
%%%%%%%%%%%%%%%%%%%%%%%%%%%%%%%%%%%%%%%%%%%%%%%%%%%%%%%%%%%%%%%%
\section{Introduction}

The Large Area Telescope \citep[LAT, ][]{atwood09} on board the {\it Fermi Gamma-ray Space Telescope} has revolutionized our understanding of the high-energy sky. The latest release of the broadband all-sky LAT catalog \citep[\ie the Third Catalog of \lat Sources or 3FGL,][]{3FGL} characterizes 3033 objects in the energy range 0.1--300~GeV from the first 4 years of LAT science data. Since the sensitivity of the instrument peaks at about 1~GeV, the 3FGL necessarily favors sources that are brightest in the GeV energy range. %This catalog represents an improvement that is over an order of magnitude in number of detected sources relative to its predecessor \citep[the third EGRET catalog of high energy sources,][which contains 271 sources]{egret99}.

The \lat collaboration has also released two hard-source catalogs that were produced using analyses optimized for energies greater than tens of GeV. The First Catalog of Hard \lat Sources \citep[1FHL,][]{1FHL} describes 514 sources detected above 10~GeV from the first 3 years of LAT data. Additionally, the Second Catalog of Hard \lat Sources \citep[2FHL,][]{2FHL} reports the properties of 360 sources detected above 50~GeV from the first 80 months of data. The 2FHL was the first LAT catalog to take advantage of the latest event-level analysis (Pass~8), which provides significant improvements in event reconstruction and classification. Pass~8 increases the sensitivity, improves the angular resolution, and also extends the useful energy range of the instrument up to 2~TeV \citep{atwood13b}. The 2FHL was intended to close the energy gap between previous \lat catalogs and the range of the current generation of Imaging Atmospheric Cherenkov Telescopes (IACTs).

Besides serving as references for works on individual sources \citep[\eg][]{aleksic14}, these LAT hard-source catalogs have been instrumental in providing promising candidates for the detection by IACTs \citep[\eg][]{abeysekara15}, enabling the search for plausible $\gamma$-ray counterparts of IceCube high-energy neutrinos \citep[\eg][]{aartsen16,padovani16}, triggering studies on unidentified sources \citep{domainko14}, and enabling new studies on the extragalactic background light \citep{dominguez15}, which yielded constraints on the  extragalactic $\gamma$-ray background \citep{broderick14,ackermann16}, and on the proton component of ultra-high energy cosmic rays \citep{berezinsky16}.

The Third Catalog of Hard \lat Sources (3FHL) is the latest addition to \lat catalogs and reports on sources detected at energies above 10~GeV. The 3FHL is constructed from the first 7 years of data and takes full advantage of the improvements provided by Pass~8 by using the point-spread function (PSF)-type event classification\footnote{A measure of the quality of the direction reconstruction is used to assign events to four quartiles.}, which improves the sensitivity.

In this work, we do not look for new extended sources but explicitly model as spatially extended sources previously resolved by the LAT along with those recently found by \citet{LAT17_10GeVES}. Given that the Cherenkov Telescope Array (CTA) is expected to have an energy threshold below 50~GeV \citep{acharya13}, the 3FHL catalog offers an excellent opportunity to relate observations from space and those that will be possible in the near future from the ground.

This paper is organized as follows: \S\ref{sec:analysis} describes the methodology used to detect sources in the LAT data and to associate these sources with known astrophysical objects at other energies. Then, \S\ref{sec:results} gives details on the structure of the 3FHL catalog and describes its main properties in the Galactic and extragalactic sky, and gives details on the newly discovered $\gamma$-ray sources. In \S\ref{sec:results}, we also discuss flux variability. Finally, we conclude in \S\ref{sec:summary}.

%%%%%%%%%%%%%%%%%%%%%%%%%%%%%%%%%%%%%%%%%%%%%%%%%%%%%%%%%%%%%%%%
%
%         Analysis 
%
%%%%%%%%%%%%%%%%%%%%%%%%%%%%%%%%%%%%%%%%%%%%%%%%%%%%%%%%%%%%%%%%
\section{Analysis}
\label{sec:analysis}

In this Section, we present our methodology for extracting the high-level information provided in the catalog from the $\gamma$-ray event-level \lat data.

\subsection{Data Selection and Software}
\label{catalog_data}

The current version of the \lat data is Pass 8 \citep{atwood13b}.
For this study we have selected Source class events in the energy range from 10~GeV to 2~TeV. Adopting a 10~GeV threshold, as was done in the 1FHL catalog \citep{1FHL}, provides the benefits of a narrow PSF, with per-photon angular resolution from $0\fdg15$ at 10 GeV to less than $0\fdg1$ above 35~GeV (68\% containment radius averaged over all event types)\footnote{\url{http://www.slac.stanford.edu/exp/glast/groups/canda/lat\_Performance.htm}}, ensuring minimal confusion and low background at the PSF scale. In that range the sensitivity of the LAT observations is limited by statistics only. We used the PSF event types appropriate for each event, to obtain the best source localizations.

We analyzed seven years of data, from 2008 August 4 to 2015 August 2 ({\it Fermi} mission elapsed time 239,557,417 to 460,250,000~s). Besides the $\sim$16\% of real time lost when passing through the South Atlantic Anomaly and in other interruptions, we have excised small intervals around bright GRBs, Solar flares, as well as bad data, resulting in 182,870,410~s (5.8 years) of good time intervals. To limit contamination from the $\gamma$-ray bright Earth limb, we enforced a selection on zenith angle ($< 105\degr$) and applied a very weak constraint on rocking angle ($< 90\degr$). The scanning mode results in maximum exposure near the north celestial pole ($4.3 \times 10^7$ m$^2$ s at 10~GeV) and minimum exposure on the celestial equator ($2.6 \times 10^7$ m$^2$ s).

\begin{figure*}[!ht]
    \centering
    \includegraphics[width=16cm,trim=0 0 0 0cm]{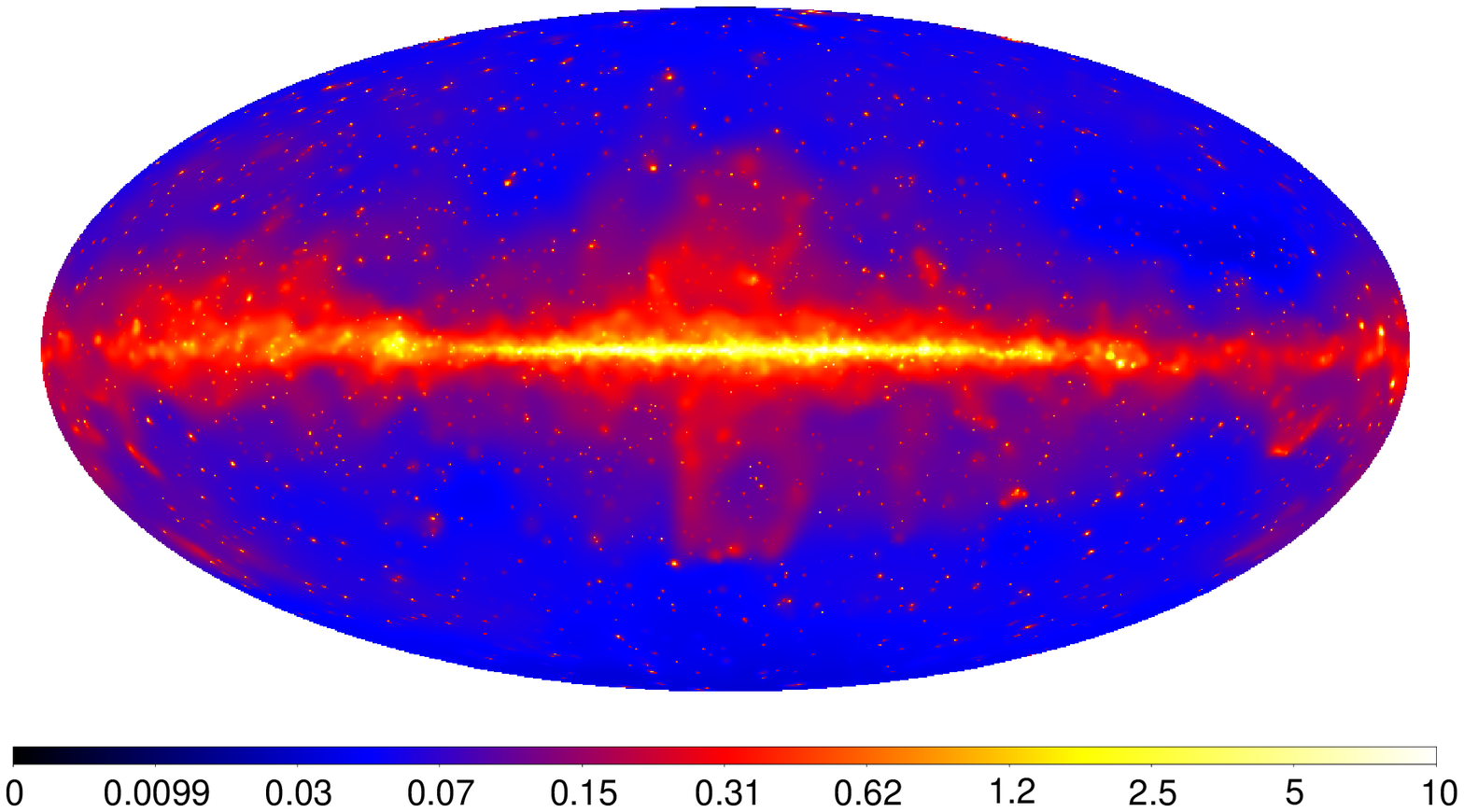} 
    \caption{Adaptively smoothed \lat counts map in the 10 GeV--2 TeV band represented in Galactic coordinates and Hammer-Aitoff projection. The image has been smoothed with a Gaussian kernel whose size was varied to achieve a minimum signal-to-noise ratio under the kernel of 2.3. The color scale is logarithmic and the units are counts per (0.1~deg)$^{2}$ pixel.
    \label{fig:counts}}
\end{figure*}

The analyzed data contain 699,582 photons at energies above 10~GeV. This is about a factor of 10 more photons than above 50~GeV in the 2FHL (60,978 photons) and more than 4 times the number in the 1FHL above 10~GeV (162,812 photons). Figure~\ref{fig:counts} shows the all-sky counts map, which has been smoothed.

We used the P8R2\_Source\_V6 instrument response functions.
We used the same models of Galactic diffuse emission and extragalactic isotropic emission as used in the 3FGL analysis, adapted to Pass 8 data and extrapolated (linearly in the logarithm) up to 2~TeV. They are available from the Fermi Science Support Center (FSSC) as {\tt gll\_iem\_v06.fits} (Galactic) and {\tt iso\_P8R2\_SOURCE\_V6\_v06.txt} (isotropic). We also used the same model as in 3FGL for the contributions from the $\gamma$-ray emissions of the Sun and Moon near the ecliptic (although their contribution above 10 GeV is very minor).

We undertook the LAT analysis using the standard $pyLikelihood$ framework (Python analog of $gtlike$) in the LAT Science Tools\footnote{See \url{http://fermi.gsfc.nasa.gov/ssc/data/analysis/documentation/Cicerone/}} (version v11r4). 
Throughout the text we use the Test Statistic TS = 2 $\Delta \log \mathcal{L}$ \citep{mattox96}, comparing the likelihood function $\mathcal{L}$ optimized with and without a given source, for quantifying how significantly a source emerges from the background.

\subsection{Source Detection}
\label{catalog_detection}

At the high energies considered here the width of the LAT PSF does not depend strongly on energy and the point-source detection is limited by source counts more than background, so we used image-based source detection techniques on counts maps integrated over all energies and event types.
The algorithm we used ({\it mr\_filter}) is based on wavelet analysis in the Poisson regime \citep{starck98}. We set the threshold to 2-$\sigma$ in the False Discovery Rate mode.
It returns a map of significant features on which we ran the peak-finding algorithm SExtractor \citep{SExtractor} to end up with a list of source candidates (hereafter seeds).
We also used another wavelet algorithm \citep[{\it PGWave},][]{damiani97,PGWave}, which differs in the detailed implementation and returns directly a list of seeds (the threshold was set to 3-$\sigma$).
Simulations indicated that the latter was somewhat more sensitive on a flat background but did not work as well in the Galactic plane.
We merged the two seed lists, eliminating duplicates within $0\fdg2$.

Since those methods work in Cartesian coordinates, we paved the sky with 26 projections in Galactic coordinates: 6 CAR (plate carr\'ee) projections along the Galactic plane covering Galactic latitudes ($b$) from $b=-$10$\degr$ to +10$\degr$, 6 AIT (Hammer-Aitoff) projections on each side of the plane covering $b=10\degr$ to $45\degr$ and 4 CAR projections covering $b=45\degr$ to 90$\degr$ in four quadrants around each pole. Each map is $5\degr$ larger on each side than the area from which the seeds are extracted, to avoid border effects. The pixel size was set to $0\fdg05$, comparable to the full width at half maximum of the PSF at high energy (the 68\% containment radius is greater than $0\fdg09$ up to 300~GeV).

Next we added seeds in the Galactic plane from the search for extended Galactic sources above 10~GeV \citep{LAT17_10GeVES} as well as seeds derived in preparatory work for the next general LAT source catalog over all energies. The full list comprised 3730 seeds.
Compared to the single {\it mr\_filter} method, adding seeds from those parallel studies and {\it PGWave} resulted in nearly 1000 more seeds, but only 24 ($< 2$\%) more sources in the final list of significant sources (\S~\ref{catalog_significance}).

The source density at $|b| > 10\degr$ is 0.036 sources per square degree (after TS selection in \S~\ref{catalog_significance}). Since the 68\% PSF containment radius is better than $0\fdg15$, confusion is rather limited. A standard plot showing the distribution of distance between sources \citep[such as Figure~13 of][]{3FGL} indicates that this catalog has missed about 20 ($<$ 2\%) high-latitude sources within $0\fdg4$ of another one.
In the Galactic plane the confusion is worse because many sources are extended.

\subsection{Localization}
\label{catalog_localization}

The position of each source was determined by maximizing the likelihood starting from the seed position, using {\it gtfindsrc}. 
We used {\it gtfindsrc} rather than {\it pointlike} (used in 3FGL) in order to benefit from the full power of PSF event types introduced in Pass 8. The {\it gtfindsrc} tool works in unbinned mode, automatically selecting the appropriate PSF for each event as a function of its event type and off-axis angle (the PSF broadens at large off-axis angles).
The {\it gtfindsrc} run was integrated into the main iterative procedure (\S~\ref{catalog_significance}), starting with the brightest sources. This ensures that the surrounding sources were correctly represented.
The main drawback is that {\it gtfindsrc} provides only a symmetric (circular) error radius, assuming a Gaussian distribution, not the full TS map and an ellipse as {\it pointlike} does. There is no reason to believe that this is a serious limitation. For example in 3FGL the average ratio between the two axes of the error ellipses was 1.20, so most ellipses were close to circular. At higher energies (1FHL) this ratio was even smaller, 1.12.

The systematic uncertainties associated with localization were not calibrated on 3FHL itself, but on the larger (and more precise) preliminary source list derived from an analysis over all energies greater than 100~MeV.
The absolute precision at the 95\% confidence level was found to be $0\fdg0075$ (it was $0\fdg005$ in 3FGL, but the statistical precision on localization was not good enough to constrain the absolute precision well).
The systematic factor was found to be 1.05, as in 3FGL.
We checked that the 3FHL localizations were consistent with the same values.
Consequently, we multiplied all error estimates by 1.05 and added 0\fdg0075 in 
quadrature.

\begin{changemargin}{0cm}{0cm}
%% Table: Extended sources
\begin{deluxetable}{llllcl}
\tabletypesize{\scriptsize}
\tablecaption{Extended Sources Modeled in the 3FHL Analysis
\label{tbl:extended}}
\tablewidth{0pt}
\tablehead{
\colhead{3FHL Name}&
\colhead{Extended Source}&
\colhead{Changes}&
\colhead{Spatial Form}&
\colhead{Extent [deg]}&
\colhead{Reference}
}

\startdata
 & SMC & Updated & Map & 1.5 & \citet{SMC_DM2016} \\
 & HB 3 & New & Disk & 0.8 & \citet{HB3_2016} \\
 & W 3 & New & Map & 0.6 & \citet{HB3_2016} \\
J0322.6$-$3712e & Fornax A & New & Map & 0.35 & \citet{Fornax2016} \\
J0427.2+5533e & FGES J0427.2+5533 & 2FHL J0431.2+5553e & Disk & 1.515 & \citet{LAT17_10GeVES} \\
 & HB 9 & New & Map & 1.0 & \citet{Araya2014_HB9} \\
J0500.9$-$6945e & LMC FarWest & New & Map\tablenotemark{a} & 0.9 & \citet{LMC2016} \\
 & LMC Galaxy & LMC & Map\tablenotemark{a} & 3.0 & \citet{LMC2016} \\
J0530.0$-$6900e & LMC 30DorWest & New & Map\tablenotemark{a} & 0.9 & \citet{LMC2016} \\
J0531.8$-$6639e & LMC North & New & Map\tablenotemark{a} & 0.6 & \citet{LMC2016} \\
J0537.6+2751e & FGES J0537.6+2751 & S 147 & Disk & 1.394 & \citet{LAT17_10GeVES} \\
J0617.2+2234e & IC 443 & Analytic & Gaussian & 0.27 & \citet{LAT10_IC443} \\
J0822.1$-$4253e & FGES J0822.1$-$4253 & Puppis A & Disk & 0.443 & \citet{LAT17_10GeVES} \\
J0833.1$-$4511e & Vela X & Analytic & Disk & 0.91 & \citet{LAT10_VelaX} \\
J0851.9$-$4620e & FGES J0851.9$-$4620 & Vela Jr & Disk & 0.978 & \citet{LAT17_10GeVES} \\
J1023.3$-$5747e & FGES J1023.3$-$5747 & New & Disk & 0.278 & \citet{LAT17_10GeVES} \\
J1036.3$-$5833e & FGES J1036.3$-$5833 & New & Disk & 2.465 & \citet{LAT17_10GeVES} \\
J1109.4$-$6115e & FGES J1109.4$-$6115 & 2FHL J1112.1-6101e & Disk & 1.267 & \citet{LAT17_10GeVES} \\
J1208.5$-$5243e & SNR G296.5+10.0 & New & Disk & 0.76 & \citet{LAT2016_SNRCat} \\
J1213.3$-$6240e & FGES J1213.3$-$6240 & New & Disk & 0.332 & \citet{LAT17_10GeVES} \\
J1303.0$-$6312e & HESS J1303$-$631 & Analytic & Gaussian & 0.24 & \citet{HESS05_J1303} \\
 & Centaurus A (lobes) & No change & Map & (2.5, 1.0) & \citet{LAT10_CenAlobes} \\
J1355.1$-$6420e & FGES J1355.1$-$6420 & 2FHL J1355.1-6420e & Disk & 0.405 & \citet{LAT17_10GeVES} \\
J1409.1$-$6121e & FGES J1409.1$-$6121 & New & Disk & 0.733 & \citet{LAT17_10GeVES} \\
J1420.3$-$6046e & FGES J1420.3$-$6046 & 2FHL J1419.3-6048e & Disk & 0.123 & \citet{LAT17_10GeVES} \\
J1443.0$-$6227e & RCW 86 & 2FHL J1443.2$-$6221e & Map & 0.3 & \citet{RCW86_2016} \\
J1507.9$-$6228e & FGES J1507.9$-$6228 & New & Disk & 0.362 & \citet{LAT17_10GeVES} \\
J1514.2$-$5909e & FGES J1514.2$-$5909 & MSH 15$-$52 & Disk & 0.243 & \citet{LAT17_10GeVES} \\
J1552.7$-$5611e & MSH 15$-$56 & New & Disk & 0.21 & \citet{LAT2016_SNRCat} \\
J1553.8$-$5325e & FGES J1553.8$-$5325 & New & Disk & 0.523 & \citet{LAT17_10GeVES} \\
J1615.3$-$5146e & HESS J1614$-$518 & Analytic & Disk & 0.42 & \citet{lande12} \\
J1616.2$-$5054e & HESS J1616$-$508 & Analytic & Disk & 0.32 & \citet{lande12} \\
J1631.6$-$4756e & FGES J1631.6$-$4756 & HESS J1632$-$478 & Disk & 0.256 & \citet{LAT17_10GeVES} \\
J1633.0$-$4746e & FGES J1633.0$-$4746 & HESS J1632$-$478 & Disk & 0.610 & \citet{LAT17_10GeVES} \\
J1636.3$-$4731e & FGES J1636.3$-$4731 & New & Disk & 0.139 & \citet{LAT17_10GeVES} \\
J1652.2$-$4633e & FGES J1652.2$-$4633 & New & Disk & 0.718 & \citet{LAT17_10GeVES} \\
J1655.5$-$4737e & FGES J1655.5$-$4737 & New & Disk & 0.334 & \citet{LAT17_10GeVES} \\
J1713.5$-$3945e & RX J1713.7$-$3946 & Corrected & Map & 0.56 & \citet{RXJ1713_2016} \\
J1745.8$-$3028e & FGES J1745.8$-$3028 & New & Disk & 0.528 & \citet{LAT17_10GeVES} \\
J1800.5$-$2343e & FGES J1800.5$-$2343 & W 28 & Disk & 0.638 & \citet{LAT17_10GeVES} \\
J1804.7$-$2144e & FGES J1804.7$-$2144 & W 30 & Disk & 0.378 & \citet{LAT17_10GeVES} \\
J1824.5$-$1351e & HESS J1825$-$137 & Analytic & Gaussian & 0.75 & \citet{LAT11_J1825} \\
J1834.1$-$0706e & FGES J1834.1$-$0706 & New & Disk & 0.214 & \citet{LAT17_10GeVES} \\
J1834.5$-$0846e & W 41 & Corrected & Gaussian & 0.23 & \citet{W41_2015} \\
J1836.5$-$0651e & FGES J1836.5$-$0651 & HESS J1837$-$069 & Disk & 0.535 & \citet{LAT17_10GeVES} \\
J1838.9$-$0704e & FGES J1838.9$-$0704 & HESS J1837$-$069 & Disk & 0.523 & \citet{LAT17_10GeVES} \\
J1840.9$-$0532e & HESS J1841$-$055 & No change & 2D Gaussian  & (0.62, 0.38) & \citet{aharonian08} \\
J1855.9+0121e & W 44 & No change & 2D Ring & (0.30, 0.19) & \citet{LAT10_W44} \\
J1857.7+0246e & FGES J1857.7+0246 & New & Disk & 0.613 & \citet{LAT17_10GeVES} \\
J1923.2+1408e & W 51C & No change & 2D Disk & (0.38, 0.26) & \citet{LAT09_W51C} \\
J2021.0+4031e & $\gamma$-Cygni & Analytic & Disk & 0.63 & \citet{lande12} \\
J2028.6+4110e & Cygnus X cocoon & Analytic & Gaussian & 3.0 & \citet{LAT11_CygCocoon} \\
 & HB 21 & Analytic & Disk & 1.19 & \citet{LAT13_HB21} \\
J2051.0+3040e & Cygnus Loop & No change & Ring & 1.65 & \citet{LAT11_CygnusLoop} \\
J2301.9+5855e & FGES J2301.9+5855 & New & Disk & 0.249 & \citet{LAT17_10GeVES} \\
\enddata

\tablenotetext{a}{Emissivity model.}

\tablecomments{~List of all sources that have been modeled as spatially extended. Sources without a 3FHL name did not reach the significance threshold in 3FHL. The Changes column gives the name of the source in previous catalogs in case of a change. The Extent column indicates the radius for Disk (flat disk) sources, the 68\% containment radius for Gaussian sources, the outer radius for Ring (flat annulus) sources, and an approximate radius for Map (external template) sources. The 2D shapes are elliptical; each pair of parameters $(a, b)$ represents the semi-major $(a)$ and semi-minor $(b)$ axes.}

\end{deluxetable}

\end{changemargin}

\subsection{Significance and Spectral Characterization}
\label{catalog_significance}

The framework for this stage of the analysis was inherited from the 3FGL catalog analysis pipeline \citep{3FGL}. It splits the sky into regions of interest (RoIs), each with typically half a dozen sources whose parameters are simultaneously optimized. The global best fit is reached iteratively, by including sources in the outer parts of the RoI from the neighboring RoIs at the previous step. Above 10~GeV the PSF is narrow so the cross-talk is small and the iteration converges rapidly.
The diffuse emission model had exactly one free normalization parameter per RoI (see Appendix~\ref{app:bkgd} for details).
We used unbinned likelihood with PSF event types over the full energy range, neglecting energy dispersion.
Extended sources (\S~\ref{catalog_extended}) were treated just as point sources, except for their spatial templates. Whenever possible we applied the new RadialDisk and RadialGaussian analytic spatial templates for the likelihood calculation. They are not pixelized and hence more precise than the map-based templates used in 3FGL.

Sources were modeled by default with a power-law (PL) spectrum (two free parameters, a normalization and a spectral photon index).
At the end of the iteration, we kept only sources with TS $>$ 25 with the PL model, corresponding to a significance of just over 4~$\sigma$ evaluated from the $\chi^2$ distribution with 4 degrees of freedom \citep[position and spectral parameters,][]{mattox96}. We also enforced a minimum number of model-predicted events $N_{\mathrm pred} \ge 4$ (only two sources were rejected because of this limit, and only two have $N_{\mathrm pred} < 5$).
We ended up with 1556 sources with TS $>$ 25, including 48 extended sources.

The alternative curved LogParabola (LP) spectral shape
\begin{equation}
\frac{{\rm d}N}{{\rm d}E} = K \left (\frac{E}{E_0}\right )^{-\alpha -
\beta\log(E/E_0)}
\label{eq:logparabola}
\end{equation}
was systematically tested, and adopted when {\tt Signif\_Curve} = $\sqrt{2 \ln (\mathcal{L}({\rm LP})/\mathcal{L}({\rm PL}))} > 3$, corresponding to 3-$\sigma$ evidence in favor of the curved model (the threshold was 4-$\sigma$ in 3FGL). Among 1556 sources, only 6 were found to be significantly curved at the 4-$\sigma$ level. Lowering the threshold to 3 $\sigma$ added 26 curved sources, whereas an average of 4.2 would be expected by chance. So most of the additional spectral curvatures between 3~$\sigma$ and 4~$\sigma$ are real.
We iterated after changing a spectral shape or removing a source.
Only 2\% of the 3FHL sources were considered significantly curved. This does not mean that sources are less curved than over the full \lat range (100~MeV--300~GeV), but only that it is more difficult to measure curvature over a restricted energy range and with limited statistics ($>$10~GeV). One of those 32 has upward curvature. This source is associated with the pulsar PSR J1418$-$6058, and that curvature marks the transition between the pulsar emission at lower $\gamma$-ray energies seen by \lat \citep{2PC} and the very high energy $\gamma$-rays from the pulsar wind nebula detected by H.E.S.S. \citep{AharonianKook06}.

Photon and energy fluxes in the 10~GeV--1~TeV band were obtained from the best spectral model. We chose to report fluxes up to 1~TeV because integrating the energy flux up to 2~TeV has larger uncertainty when the photon index is harder than 2. Uncertainties were obtained by linear error propagation from the original parameters. No systematic errors were included. Fluxes in five energy bands were extracted in the same way as in 3FGL. The energy limits were set to 10, 20, 50, 150, 500~GeV and 2~TeV. The width of the energy bins (in the logarithm) increases with energy in order to partially compensate for the decrease of photons due to the falling source spectra. Systematic uncertainties are estimated to be 5\% in the first three bands, then 9\% and 15\%\footnote{\url{http://fermi.gsfc.nasa.gov/ssc/data/analysis/LAT\_caveats.html}}. They are not included in the five individual uncertainties.

As for 2FHL, we evaluated the probabilities that the photons near each source originated from the source using {\it gtsrcprob}, and found the highest-energy photon with probability $>$ 85\%.
%The probability of each event to belong to a source was computed using {\it gtsrcprob} and the highest-energy photon with probability $>$ 85\% was found, following what was done in 2FHL.

\subsection{Extended Sources}
\label{catalog_extended}

This work does not involve looking for new extended sources, or testing possible extension of sources detected as point-like. As in the 3FGL catalog, we explicitly modeled as spatially extended those sources that had been shown in dedicated analyses to be resolved by the LAT\footnote{The templates and spectral models are available through the {\it Fermi} Science Support Center.}. The spectral parameters of each extended source were fitted in the same way as those of point sources. We did not attempt to refit the spatial shapes.
Because many of those extended sources are much broader than the PSF at 10~GeV, we allowed adding new seeds inside the extended sources when their radii were larger than 0\fdg4 (this differs from what was done at lower energies in 3FGL). Identified point sources (in practice, pulsars) were allowed in extended sources of any size.

The 3FGL catalog considered 25 extended sources. Five more were introduced in the 2FHL catalog. Several descriptions of extended sources were improved upon since then (see Table~\ref{tbl:extended} for details). In particular the Large Magellanic Cloud (LMC) is now represented by four independent components (but only the hard compact components are detected above 10~GeV, not the large one at the scale of the entire galaxy). The RCW 86 and RX J1713.7$-$3946 supernova remnants (SNRs) also benefited from new templates. The W~41 template was corrected (an error made 2FHL J1834.5$-$0846e too narrow). A new extragalactic extended source was reported in the lobes of the Fornax A radio galaxy. Two (G296.5+10.0 and MSH 15$-$56 = G326.3$-$1.8) were taken from the systematic study of Galactic SNRs.

A recent comprehensive search for extended sources above 10 GeV in the Galactic plane \citep[$|b|<7^{\circ}$, ][]{LAT17_10GeVES} resulted in 46 detections, all represented as disks. Eleven of those are new, and were entered as such in 3FHL (sources numbered FGES J1745.8$-$3028, J1857.7+0246, J2301.9+5855, J1023.3$-$5747, J1036.3$-$5833, J1213.3$-$6240, J1409.1$-$6121, J1507.9$-$6228, J1553.8$-$5325, J1652.2$-$4633, J1655.5$-$4737). FGES J1745.8$-$3028 was flagged in their work because the disk size was unstable with respect to the underlying diffuse model. We introduced it anyway, because it appears very significant (TS = 114) and preferred to two point sources. The 35 other detections coincide with previously detected extended sources. We switched to the new description only when it was clearly warranted, \ie the fit was improved by $\Delta \log \mathcal{L} > 15 + p$ where $p$ is the number of additional parameters as in the Akaike criterion \citep{akaike74}. We also kept the previous Gaussian template of the IC 443 SNR because it is closer to the radio SNR than the two disks in the FGES representation and allows collecting all the flux into a single source, even though it exceeds the above criterion. Only one Galactic source not detected in FGES appears in 3FHL. This is the Cygnus Loop, which has TS just above threshold, and can be easily understood because \citet{LAT17_10GeVES} analyzed only six years of data instead of seven here.

Thirteen extended source templates were abandoned in favor of the fitted disk representations:
\begin{itemize}
\item W~28 is represented by the broader disk FGES J1800.5$-$2343 which encompasses the four sources found outside the SNR \citep{W28_2014}. Three of those sources were too faint to be recovered individually by the point source detection algorithm. The brightest peak in W~28 proper, as well as the brightest outer peak (HESS J1800$-$240 B) were detected as individual point sources on top of FGES J1800.5$-$2343.
\item W~30 is represented by the disk FGES J1804.7$-$2144, shifted by about 0\fdg2 with respect to the 3FGL disk. There is no doubt that the emission around 1~GeV, which is close to the SNR \citep{LAT12_W30}, is not centered on the same direction as the emission above 10~GeV, which is closer to the TeV source HESS J1804$-$216 \citep{aharonian06_gps}.
\item Two sources within one degree of each other, G24.7+0.6 and HESS J1837$-$069, came from previous similar automatic searches for extended sources \citep[][respectively]{LAT2016_SNRCat,lande12}. They were replaced by the better representation involving three overlapping disks FGES J1834.1$-$0706, J1836.5$-$0651, J1838.9$-$0704.
\item The previous templates for 2FHL sources SNR G150.3+4.5, J1112.1$-$6101e, HESS J1356$-$645 and J1420$-$607 \citep{2FHL} were replaced by the disks FGES J0427.2+5533, J1109.4$-$6115, J1355.1$-$6420 and J1420.3$-$6046 (with much better statistics down to 10~GeV). FGES J0427.2+5533 is actually closer in size to the radio SNR (larger than the 2FHL size). FGES J1355.1$-$6420 and J1420.3$-$6046 are smaller than their 2FHL counterparts, closer to the TeV size.
\item The radio template for S 147 \citep{LAT12_S147} was replaced by the flat disk FGES J0537.6+2751. With the flat disk that source is significant, whereas the radio template resulted in TS $<$ 25.
\item The previous template for Puppis A \citep{2012LAT_PuppisA} was replaced by the somewhat broader disk FGES J0822.1$-$4253 which follows more closely the radio and X-ray contour. Without this change a point source was necessary to fit the data just outside the previous disk.
\item The previous template for Vela Jr \citep{LAT11_RXJ0852} was replaced by the somewhat smaller disk FGES J0851.9$-$4620, closer in size to the X-ray SNR.
\item The previous template for the pulsar wind nebula (PWN) MSH 15$-$52 \citep{LAT10_PSR1509} was replaced by FGES J1514.2$-$5909, shifted by about 0\fdg1 with respect to the 3FGL disk. Without this shift a point source close to PSR B1509$-$58 was necessary, although this pulsar is known to have a very soft spectrum.
\item The previous template for HESS J1632$-$478 \citep{lande12} was replaced by the combination of a broader disk (FGES J1631.6$-$4756) with two smaller ones on top of it (FGES J1633.0$-$4746 and J1636.3$-$4731), all with different spectra. Together they provide a much better representation at the cost of only two additional parameters (for extension) since two point sources were necessary next to HESS J1632$-$478.
\end{itemize}

For each of the extended sources, Table~\ref{tbl:extended} lists its name, changes since 3FGL and 2FHL if any, the spatial template description, the extent and the reference to the dedicated analysis. In the catalog these sources are tabulated with the point sources, with the only distinction being that no position uncertainties are reported and their names end in \texttt{e} (see \S~\ref{sec:catalog}). Point sources inside extended ones are marked by ``xxx field'' in the \texttt{ASSOC2} column of the catalog.

\subsection{Background-only Simulation}
\label{catalog_simus}

The narrow PSF above 10~GeV implies that the number of independent positions in the sky is large. Therefore we might expect a fraction of spurious sources from background fluctuations only. Taking a disk of radius $0\fdg15$ (68\% PSF containment radius at 10~GeV) as the source size, there are $6 \times 10^5$ independent positions in the sky. Since the probability to reach TS $>$ 25 ($\chi^2$ distribution with 4 degrees of freedom) is $5 \times 10^{-5}$, we might expect as many as 30 spurious sources.

In order to quantify this more precisely we simulated the full sky with the same exposure as the real data, assuming pure (Galactic + isotropic) background. Owing to the narrowness of the PSF above 10~GeV there is essentially no correlation between sources so it is not worth including sources in the simulation.
The source detection step was not run in exactly the same way as on the real data, because {\it mr\_filter} uses a False Detection Rate threshold, which depends on the number of true sources. Since there is no true source in the simulation, the same setting would have resulted in very few seeds at the detection step. Instead we used a flat threshold at 4-$\sigma$ over all wavelet scales, resulting in 135 seeds over the entire sky.
% This is more than enough {\bf to feed the likelihood step (the lowest significance among those which did reach TS$> 25$ was 4.6 $\sigma$, ranked 47 in order of decreasing {\it mr\_filter} significance)}.
We merged those with the PGWave seeds which are much more numerous (1955) because of the 3-$\sigma$ threshold.

The maximum likelihood analysis of these seeds (including localization) resulted in ten sources at TS $>$ 25 including two at TS $>$ 30, randomly distributed over the sky, with $N_{\mathrm pred} \simeq 6$ and formal position error $\sim 0\fdg05$ (similar to the faint sources in the real data). We conclude that the number of spurious sources in 3FHL is closer to ten than the rough estimate of 30 given at the beginning of this section.
This is a small fraction even of sources close to threshold in the real sky (192 at 25 $\le$ TS $<$ 30 and 161 at 30 $\le$ TS $<$ 35).

%%%%%%%%%%%%%%%%%%%%%%%%%%%%%%%%%%%%%%%%%%%%%%%%%%%%%%%%%%%%%%%%
%
%         Associations
%
%%%%%%%%%%%%%%%%%%%%%%%%%%%%%%%%%%%%%%%%%%%%%%%%%%%%%%%%%%%%%%%%
\subsection{Source Association and Classification}
\label{sec:associations}
We adopt the same procedure for evaluating the probabilities of association between $\gamma$-ray sources and potential counterparts previously used in 3FGL. This procedure is based on two different association methods, the Bayesian method  \citep{1FGL} and the the Likelihood-Ratio (LR) method \citep{2LAC,3LAC}. The fractions of new associations provided by the two methods are different from 3FGL since the source populations are different, as described in \S\ref{sec:ext}. Following the same strategy as in the 3FGL, we distinguish between associated and identified sources. Associations depend primarily on close positional correspondence whereas identifications require measurement of correlated variability at other wavelengths or characterization of the 3FHL source by its angular extent.

The Bayesian method is applied using the set of potential-counterpart catalogs listed in Table 12 of \cite{3FGL}, updated to the latest available versions. The priors are recalibrated via Monte-Carlo simulations to enable a proper estimate of the association probabilities and in turn of the false association rates. These rates indeed depend on the sizes of the error ellipses of the sources, whose distributions are appreciably  different in the 3FGL and 3FHL catalogs.  A total of 1187 associations with posterior probabilities greater than 0.80 are found via this method, with an estimated number of false positives of $\sim$5. Thanks to the updated catalogs \citep[\eg][Pe\~na-Herazo et al. in preparation]{bzcat5,alvarezcrespo16a,alvarezcrespo16b} 25 unidentified 3FGL sources detected in 3FHL are now associated with blazars.

\begin{figure*}[!ht]
    \centering
    \includegraphics[width=8cm,trim=0.4cm 0.2cm 1.8cm 1.4cm,clip=True]{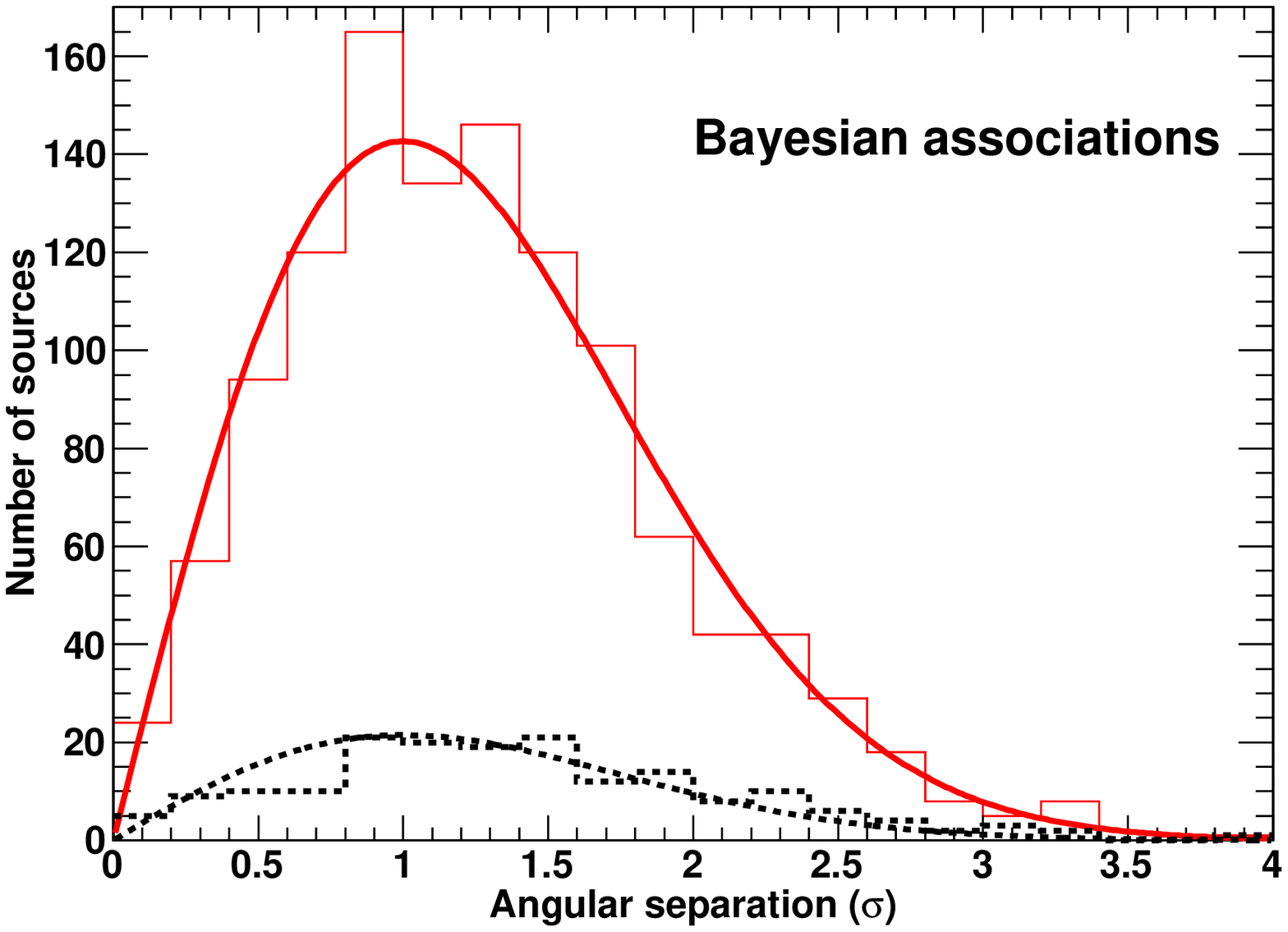}
    \includegraphics[width=8cm,trim=0.4cm 0.2cm 1.8cm 1.4cm,clip=True]{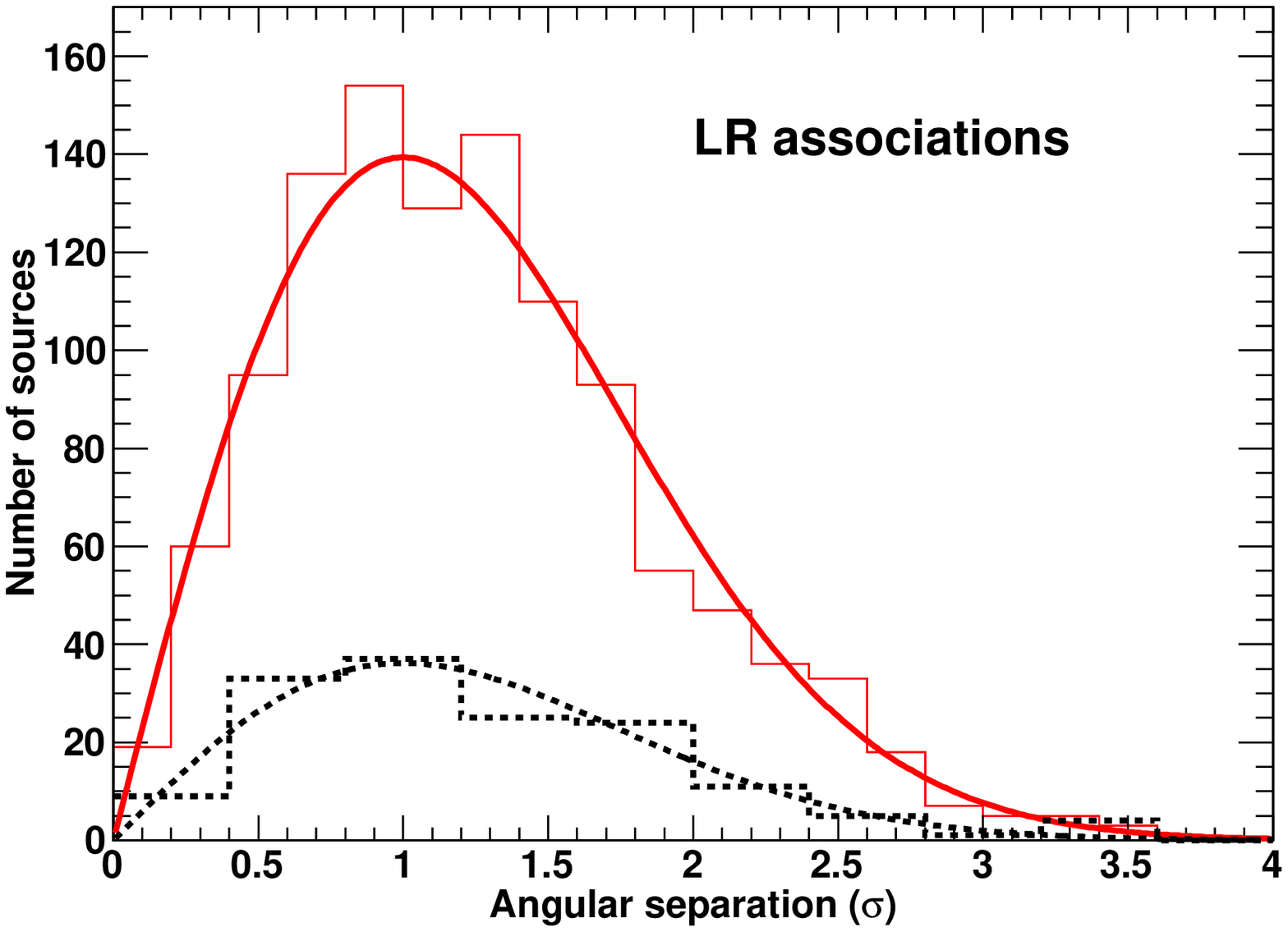}
    \caption{Distributions of angular separations in $\sigma$ units between 3FHL sources and their counterparts ($r95=2.448\sigma$). {\it (Left panel)}: Sources associated with the Bayesian method (red solid line) and sources solely associated with that method (black dotted line). {\it (Right panel)}: Same, but for the LR method. The curves correspond to the expected distributions for real associations.
    \label{fig:angsep}}
\end{figure*}

The Likelihood-Ratio method  provides additional associations with blazar candidates based on large radio and X-ray surveys (typically including $> 10^5$ sources). The Bayesian method in its current implementation cannot handle these surveys since their large source densities conduce to too many false positives.  The resulting associated counterparts are then scrutinized using additional available multi-wavelength data to assess their classification. If no or too limited information is found,  a sole association with radio counterparts is usually rejected, the high source densities in the radio surveys making the chance of false positives exceedingly large.   A total of 1151 3FHL sources are associated by this method, of which 150 (with an estimated number of 15 false positives) are not associated by the Bayesian method. In these 150 sources, 44 have 3FGL counterparts as well.  As an exception to the rejection criterion outlined above, if the counterpart belongs to the ROSAT X-ray survey (which has a  much lower source density than the radio surveys), we do report the association with a classification left as {\it unknown}. These sources (23 in total) are not  particularly different from the unassociated sources in their $\gamma$-ray properties or sky locations. Follow-up observations would be particularly useful to determine their nature. Associations with $\gamma$-ray sources reported in earlier LAT catalogs are established by requiring an overlap of their respective 95\%  error ellipses.  Note that in the rare cases (6 in total) where conflicting Bayesian-based associations are found for a 3FHL source and its 3FGL counterpart, we give precedence to the choice presenting the smaller error ellipse, foregoing consistency with 3FGL in some cases. The results of the association procedures are summarized in Table~\ref{tab:classes}. Figure~\ref{fig:angsep}  shows the distributions of angular separation between the 3FHL sources and their assigned counterparts. The good agreement with  the expected distributions for real associations visible in this figure provides confidence in the reliability of the associations.

The associated blazars were optically classified as flat-spectrum radio quasars (FSRQs), BL Lacs and blazars of unknown types \citep[BCUs, ][]{shaw12,shaw13}. The source peak frequencies were adopted from 3LAC \citep{3LAC} when available or determined via the same approach. Low-synchrotron peak (LSP), intermediate-synchrotron peak (ISP), and high-synchrotron peak (HSP) blazars are those with $\log_{10}(\nu^{s}_{peak})<14$, $14<\log_{10}(\nu^{s}_{peak})<15$, $\log_{10}(\nu^{s}_{peak})>15$, respectively, with $\nu^{s}_{peak}$ given in units of Hz.

%%%%%%%%%%%%%%%%%%%%%%%%%%%%%%%%%%%%%%%%%%%%%%%%%%%%%%%%%%%%%%%%
%
%         Results
%
%%%%%%%%%%%%%%%%%%%%%%%%%%%%%%%%%%%%%%%%%%%%%%%%%%%%%%%%%%%%%%%%
\section{The 3FHL Catalog}
\label{sec:results}

\begin{figure*}[!ht]
    \centering
    \includegraphics[width=16cm,trim=0 0 0 0cm]{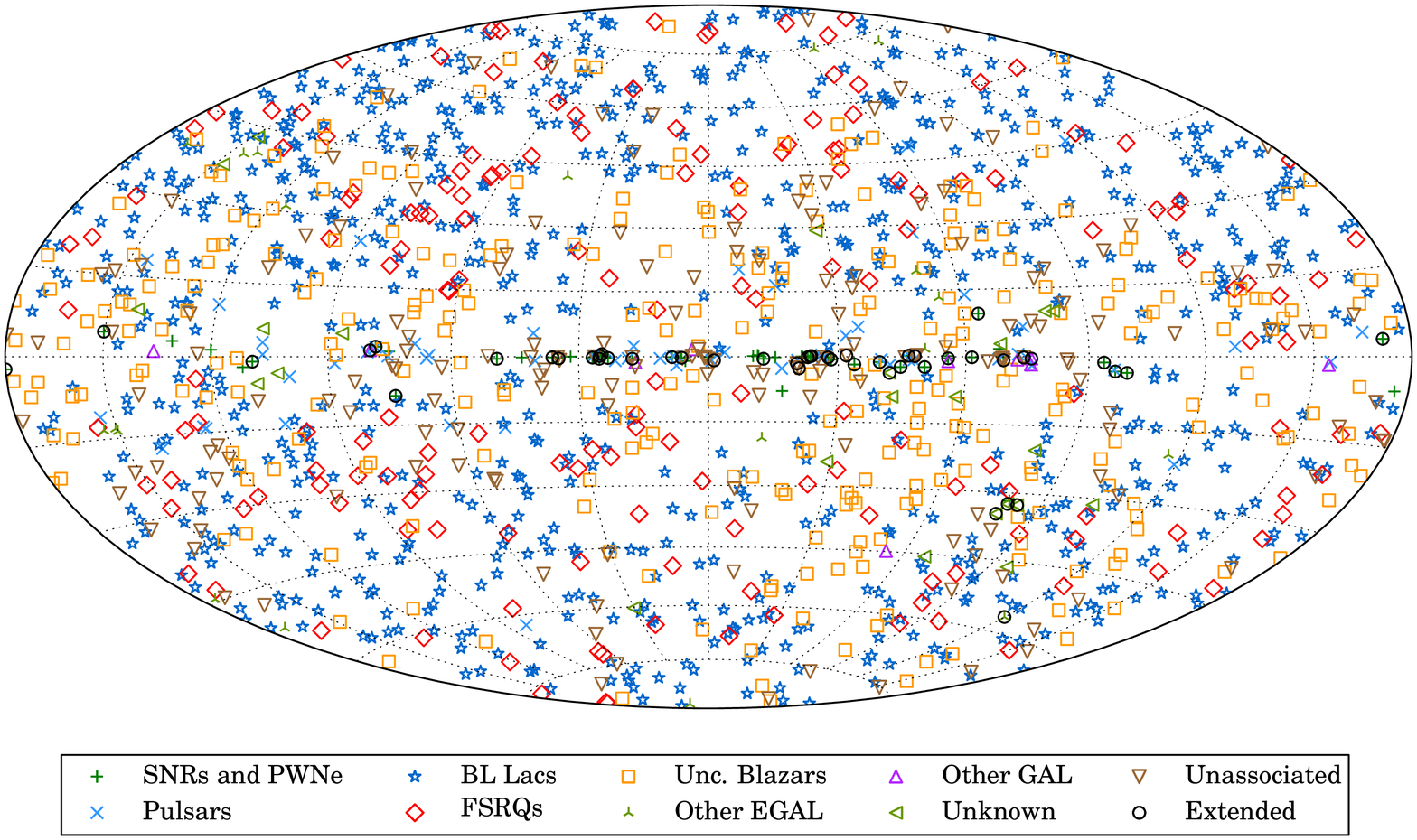} 
    \caption{Sky map, in Galactic coordinates and Hammer-Aitoff projection,
showing the objects in the 3FHL catalog classified by their most likely source classes. 
    \label{fig:all_sky}}
\end{figure*}

The 3FHL catalog includes 1556 sources detected over the whole sky. \citep[We note that the number of sources in the 3FHL catalog is more than the 1506 $\gamma$ rays detected above 10~GeV by the EGRET experiment on the predecessor {\it Compton} Gamma-ray Observatory mission, ][]{thompson05}. The association procedure (see \S\ref{sec:associations}) finds that 79\% of the sources in the catalog (1231 sources) are extragalactic, 8\% (125) are Galactic, and 13\% (200) are unassociated (or associated with a source of unknown nature). Of the unassociated/unknown sources, 83 are located at $|b|<10^{\circ}$, and 117 at $|b|\geq 10^{\circ}$. Since sources outside the plane are typically extragalactic, the fraction of extragalactic sources in the sample is likely about 87\%. Figure~\ref{fig:all_sky} shows the locations of 3FHL sources color-coded by source class.

% Table listing the source classes and their numbers
\begin{deluxetable}{lcrcr}
\setlength{\tabcolsep}{0.04in}
\tablewidth{0pt}
\tabletypesize{\small}
\tablecaption{3FHL Source Classes \label{tab:classes}}
\tablehead{
\colhead{Description} & 
\multicolumn{2}{c}{Identified} &
\multicolumn{2}{c}{Associated} \\
& 
\colhead{Designator} &
\colhead{Number} &
\colhead{Designator} &
\colhead{Number}
}
\startdata
Pulsar & PSR & 53 & psr & 6 \\
Pulsar Wind Nebula & PWN & 9 & pwn & 8\\
Supernova remnant & SNR & 13 & snr & 17 \\
Supernova remnant / Pulsar wind nebula &  \nodata  &  \nodata  & spp & 9 \\
High-mass binary & HMB & 4 & hmb & 1 \\
Binary & BIN & 1 &  \nodata  &  \nodata \\
Globular cluster & \nodata & \nodata &  glc  &  2 \\
Star-forming region & SFR & 1 &  sfr  &  1  \\
\hline
Starburst galaxy &  \nodata  &  \nodata  & sbg & 4 \\
BL Lac type of blazar & BLL & 19 & bll & 731 \\
Flat spectrum radio quasar type of blazar & FSRQ & 30 & fsrq & 142 \\
Non-blazar active galaxy &  \nodata  &  \nodata & agn & 1 \\
Narrow-line seyfert 1 & NYLS1 & 1 &  \nodata  &  \nodata  \\
Radio galaxy & RDG & 4 & rdg & 9 \\
Blazar candidate of uncertain type &  \nodata  &  \nodata  & bcu & 290 \\
\hline
Total & identified & 136 & associated & 1220 \\
\hline
Unclassified &   \nodata  &   \nodata  & unknown & 23 \\
Unassociated & \nodata & \nodata &  \nodata & 177 \\
Total in the 3FHL & \nodata & \nodata & \nodata & 1556 \\
\enddata
\tablecomments{ The designation `spp' indicates potential association with SNR or PWN. Designations shown in capital letters are firm identifications; small letters indicate associations. Note: The PWN N 157 B in the LMC is counted as Galactic.}
\end{deluxetable}

%%%%%%%%%%%%%%%%%%%%%%%%%%%%%%%%%%%%%%%%%%%%%%%%%%%%%%%%%%%%%%%%
%
%         Description of the Catalog
%
%%%%%%%%%%%%%%%%%%%%%%%%%%%%%%%%%%%%%%%%%%%%%%%%%%%%%%%%%%%%%%%%
\subsection{Description of the Catalog}
\label{sec:catalog}

The FITS format of the 3FHL catalog\footnote{The file is available  from the $Fermi$ Science Support Center.} is similar to that of the 3FGL catalog \citep{3FGL}. The file has four binary table extensions. The {\tt LAT\_Point\_Source\_Catalog} extension has all of the information about the sources (see Table~\ref{tab:description} for details). The catalog is available in a .tar.gz package.% \sout{For reference, an excerpt of the catalog itself is shown in Table~\ref{tab:catalog}}.

Relative to previous LAT catalogs, two changes are important:
\begin{itemize}
\item The parameters of the curved (LogParabola) spectral shape, which is systematically tested against a power law, are now always reported via the {\tt Spectral\_Index} and {\tt beta} columns, even when the curvature is not significant ({\tt Signif\_Curve} $<$ 3). The photon index of the power-law model is always reported via the {\tt PowerLaw\_Index} column. In 3FGL {\tt Spectral\_Index} contained the power-law index when the power-law model was adopted. The {\tt Flux\_Density}, {\tt Flux} and {\tt Energy\_Flux} columns still refer to the preferred model ({\tt SpectrumType}).
\item The format of the spectral energy distributions (SEDs) differs. Now, we give the fluxes and their uncertainties for each source in a vector column matching the number of energy bins. These bins are documented in the {\tt EnergyBounds} extension. The level of the relative systematic uncertainty on the effective area in each band ({\tt SysRel} column) is given in the same extension.
\end{itemize}

The extensions {\tt ExtendedSources} and {\tt ROIs} (format unchanged since 3FGL) contain information about the 55 extended sources (Table~\ref{tbl:extended}) that were included in the analysis (only 48 were detected) and the 741 ROIs over which the analysis ran. The extended sources are singled out by an {\tt e} appended to their names in the main table. The background parameters are reported in the {\tt ROIs} extension following the model described in App.~\ref{app:bkgd}. The {\tt GTI} extension is not included because it would dominate the volume of the file.

% Table of columns in FITS version
\begin{deluxetable}{lccl}
\setlength{\tabcolsep}{0.04in}
\tablewidth{0pt}
\tabletypesize{\scriptsize}
\tablecaption{LAT 3FHL FITS Format: LAT\_Point\_Source\_Catalog Extension\label{tab:description}}
\tablehead{
\colhead{Column} &
\colhead{Format} &
\colhead{Unit} &
\colhead{Description}
}
\startdata
Source\_Name & 18A & \nodata & Official source name 3FHL JHHMM.m+DDMM \\
RAJ2000 & E & deg & Right Ascension \\
DEJ2000 & E & deg & Declination \\
GLON & E & deg & Galactic Longitude \\
GLAT & E & deg & Galactic Latitude \\
Conf\_95\_SemiMajor & E & deg & Error radius at 95\% confidence \\
Conf\_95\_SemiMinor & E & deg & = Conf\_95\_SemiMajor in 3FHL \\
Conf\_95\_PosAng & E & deg & NULL in 3FHL (error circles) \\
ROI\_num & I & \nodata & ROI number (cross-reference to ROIs extension) \\
Signif\_Avg & E & \nodata & Source significance in $\sigma$ units over the 10~GeV to 2~TeV band \\
Pivot\_Energy & E & GeV & Energy at which error on differential flux is minimal \\
Flux\_Density & E & cm$^{-2}$ GeV$^{-1}$ s$^{-1}$ & Differential flux at Pivot\_Energy \\
Unc\_Flux\_Density & E & cm$^{-2}$ GeV$^{-1}$ s$^{-1}$ & $1\sigma$  error on differential flux at Pivot\_Energy \\
Flux & E & cm$^{-2}$ s$^{-1}$ & Integral photon flux from 10~GeV to 1~TeV obtained by spectral fitting \\
Unc\_Flux & E & cm$^{-2}$ s$^{-1}$ & $1\sigma$ error on integral photon flux from 10~GeV to 1~TeV \\
Energy\_Flux & E & erg cm$^{-2}$ s$^{-1}$ & Energy flux from 10~GeV to 1~TeV obtained by spectral fitting \\
Unc\_Energy\_Flux & E & erg cm$^{-2}$ s$^{-1}$ & $1\sigma$ error on energy flux from 10~GeV to 1~TeV \\
Signif\_Curve & E & \nodata & Significance (in $\sigma$ units) of the fit improvement between power-law and \\
& & & LogParabola. A value greater than 3 indicates significant curvature \\
SpectrumType & 18A & \nodata & Spectral type (PowerLaw or LogParabola) \\
Spectral\_Index & E & \nodata & Best-fit photon number index at Pivot\_Energy  when fitting with LogParabola \\
Unc\_Spectral\_Index & E & \nodata & $1\sigma$ error on Spectral\_Index \\
beta & E & \nodata & Curvature parameter $\beta$ when fitting with LogParabola \\
Unc\_beta & E & \nodata & $1\sigma$ error on $\beta$ \\
PowerLaw\_Index & E & \nodata & Best-fit photon number index when fitting with power law \\
Unc\_PowerLaw\_Index & E & \nodata & $1\sigma$ error on PowerLaw\_Index \\
Unc\_Flux\_Band & 10E & cm$^{-2}$ s$^{-1}$ & $1\sigma$ lower and upper error on Flux\_Band\tablenotemark{a} \\
nuFnu & E & erg cm$^{-2}$ s$^{-1}$ & Spectral energy distribution over each spectral band \\
Sqrt\_TS\_Band & E & \nodata & Square root of the Test Statistic in each spectral band \\
Npred & E & \nodata & Predicted number of events in the model \\
HEP\_Energy & E & GeV & Highest energy among events probably coming from the source \\
HEP\_Prob & E & \nodata & Probability of that event to come from the source \\
Flux\_Band & 5E & cm$^{-2}$ s$^{-1}$ & Integral photon flux in each spectral band \\
Variability\_BayesBlocks & I & \nodata & Number of Bayesian blocks from variability analysis; 1 if not variable, \\
& & &  -1 if could not be tested \\
Extended\_Source\_Name & 18A & \nodata & Cross-reference to the ExtendedSources extension \\
ASSOC\_GAM & 18A & \nodata & Correspondence to previous $\gamma$-ray source catalog\tablenotemark{b} \\
TEVCAT\_FLAG & A & \nodata & P if positional association with non-extended source in TeVCat \\
& & & E if associated with an extended source in TeVCat, N if no TeV association \\
& & & C if TeV source candidate as defined in \S~\ref{newandTeV} \\
ASSOC\_TEV & 24A & \nodata & Name of likely corresponding TeV source from TeVCat, if any \\
CLASS & 7A & \nodata & Class designation for associated source; see Table~\ref{tab:classes} \\
ASSOC1 & 26A & \nodata & Name of identified or likely associated source \\
ASSOC2 & 26A & \nodata & Alternate name or indicates whether the source is inside an extended source \\
ASSOC\_PROB\_BAY & E & \nodata & Probability of association according to the Bayesian method \\
ASSOC\_PROB\_LR & E & \nodata & Probability of association according to the Likelihood Ratio method \\
Redshift & E & \nodata & Redshift of counterpart, if known \\
NuPeak\_obs & E & Hz & Frequency of the synchrotron peak of counterpart, if known \\
\enddata
\tablenotetext{a} {Separate $1\sigma$ errors are computed from the likelihood profile toward lower and larger fluxes. The lower error is set equal to NULL and the upper error is derived from a Bayesian upper limit if the $1\sigma$ interval contains 0 ($TS < 1$).}
\tablenotetext{b} {in the order 3FGL $>$ 2FHL $>$ 1FHL $>$ 2FGL $>$ 1FGL $>$ EGRET.}
\end{deluxetable}

%\input{tab_catalog.tex}

%\label{sec:catalog}

%%%%%%%%%%%%%%%%%%%%%%%%%%%%%%%%%%%%%%%%%%%%%%%%%%%%%%%%%%%%%%%%
%
%         General Results
%
%%%%%%%%%%%%%%%%%%%%%%%%%%%%%%%%%%%%%%%%%%%%%%%%%%%%%%%%%%%%%%%%
\subsection{General Characteristics of Sources}
%We describe the general spectral and positional characteristics of the 3FHL sources in this section.

The 3FHL sources have integrated fluxes above 10~GeV that range from $1.3\times 10^{-11}$~ph~cm$^{-2}$~s$^{-1}$ (approximately 0.5\% of the Crab Nebula flux) to $1.2\times 10^{-8}$~ph~cm$^{-2}$~s$^{-1}$ with a median of $5.0\times 10^{-11}$~ph~cm$^{-2}$~s$^{-1}$.

\begin{figure}[!ht]
    \centering
    \includegraphics[width=\columnwidth]{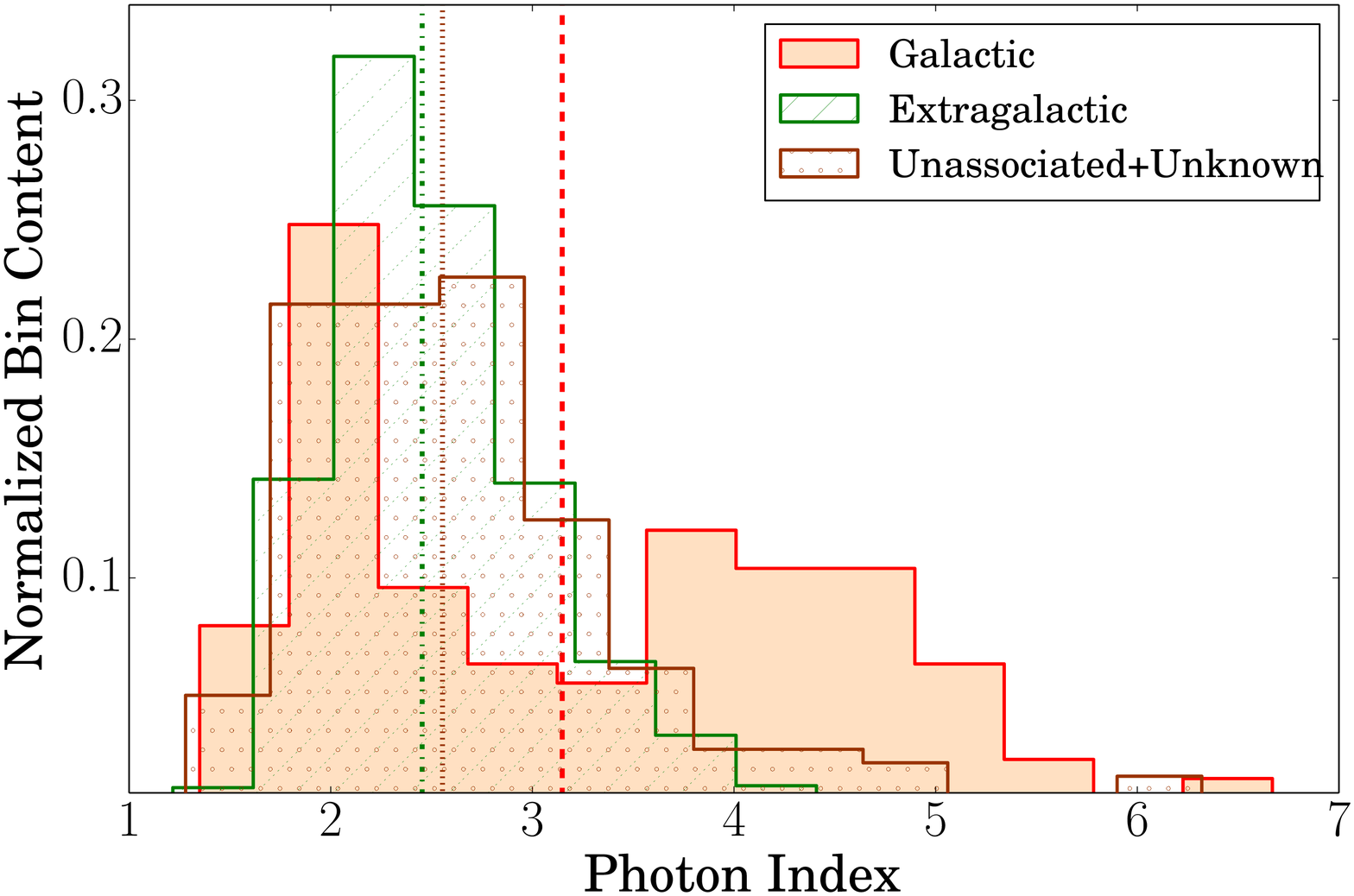} 
    \caption{Normalized distributions of the spectral indices of the Galactic sources (orange), extragalactic sources (green slash), and unassociated plus unknown sources (brown dotted) in 3FHL. The medians of the distributions are plotted with dashed, dashed-dotted, and dotted vertical lines, respectively.
    \label{fig:hist_index}}
\end{figure}

\begin{figure}[!ht]
    \centering
    \includegraphics[width=\columnwidth]{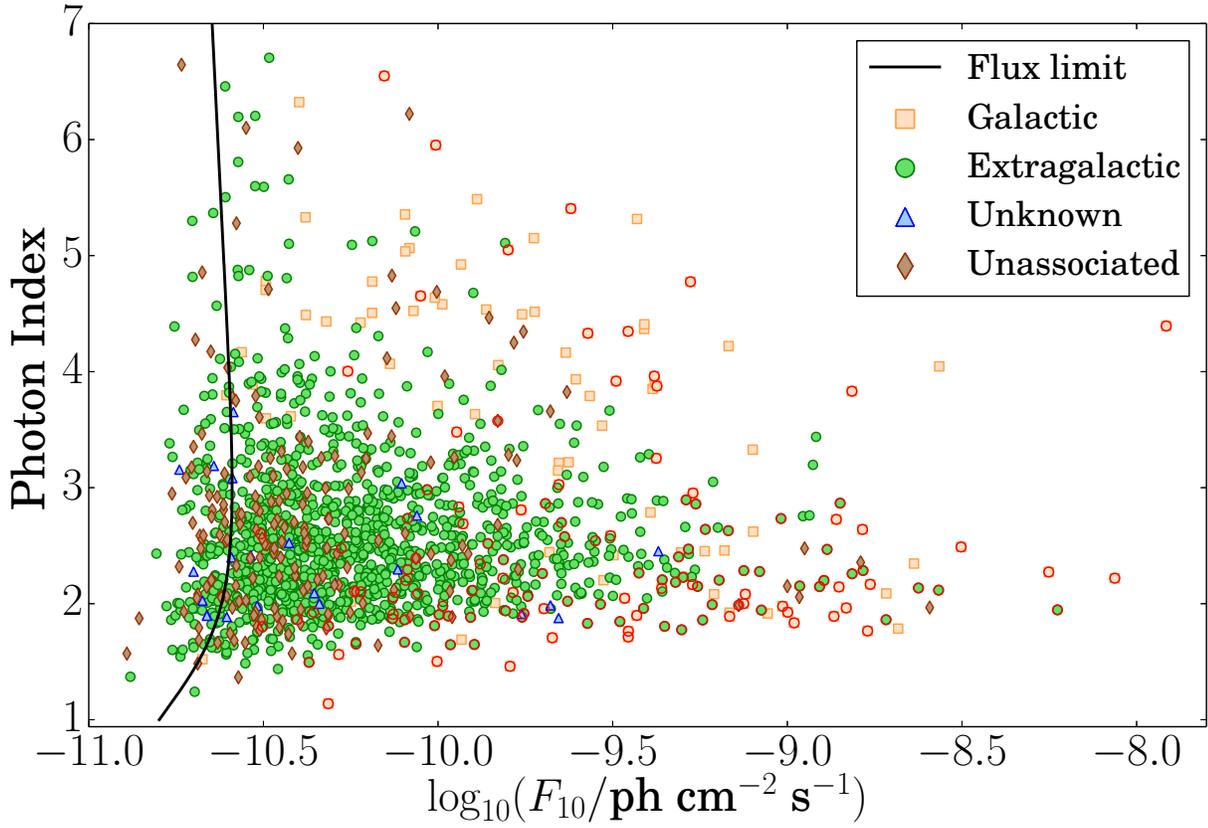} 
    \caption{The spectral index of Galactic (orange squares), extragalactic (green circles), unknown (blue triangles) and unassociated sources (brown diamonds) versus the integrated photon flux above 10~GeV. The black line shows the flux limit averaged over the high latitude sky ($|b|\geq 10^{\circ}$). Symbols outlined with red are in the TeVCat.
    \label{fig:index_vs_flux}}
\end{figure}

The median spectral index is 2.48, which is characteristic of relatively hard sources. In Figure~\ref{fig:hist_index}, we show the spectral index distributions by source class. The figure shows that Galactic sources tend to have a broader range of spectral indices whereas the distribution of extragalactic sources peaks at about an index of 2.  There is also a clear bimodality in the Galactic index distribution produced by the SNR+PWN and PSR populations (see \S\ref{sec:galactic} for details). The population of unknown sources follows a similar trend as the blazars, SNRs, and PWNe but different from PSRs. The median of the positional uncertainty is $0.038^{\circ}$ (2.3~arcmin; we note that about 75\% of the 3FGL sources present in the 3FHL now have smaller localization uncertainties). Figure~\ref{fig:index_vs_flux} illustrates that the detection threshold on photon flux does not depend strongly on spectral index (the same is not true for the energy flux). The reason for this is the constancy of the LAT per-photon resolution with energy at $E\geq 10$~GeV. We note that extragalactic sources are detected to lower fluxes than Galactic objects, highlighting that the sensitivity for source detection becomes worse in the plane of the Galaxy. Figure~\ref{fig:sensitivity} shows the flux sensitivity as a function of sky location \citep[see the appendix of ][]{1FHL}.

Figure~\ref{fig:seds} shows example SEDs for four sources over four decades in energy, which combine 3FGL and 3FHL spectral data (the 1FHL data is also shown for comparison).

\begin{figure*}[!ht]
    \centering
    \includegraphics[width=16cm,trim=0 0 0 0cm]{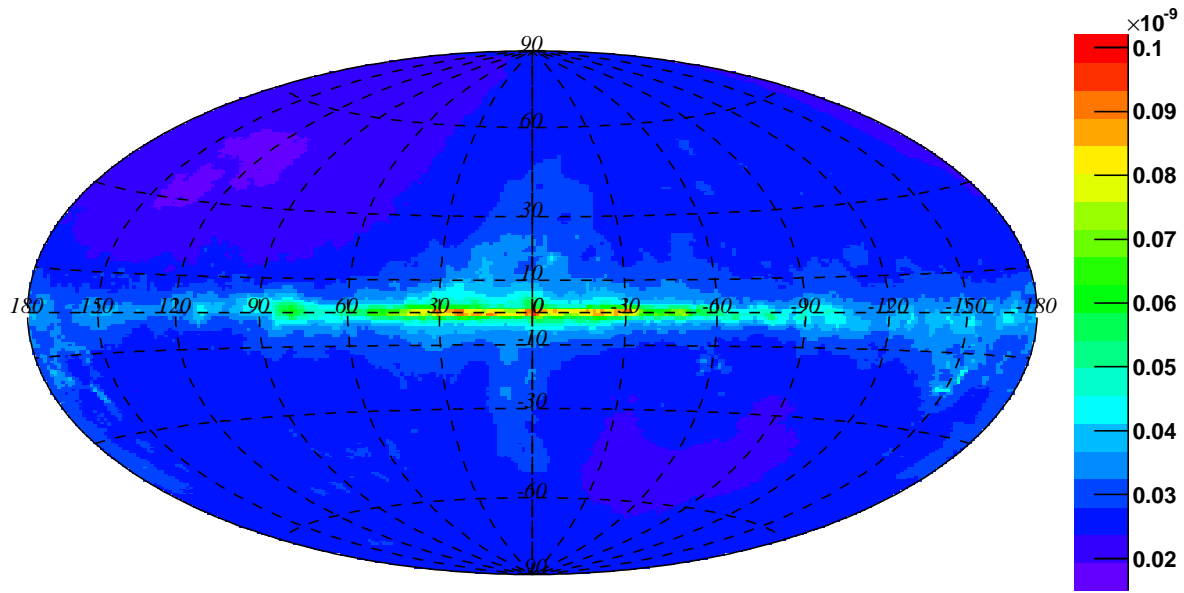}
	\caption{Point-source flux limit in units of ph~cm$^{-2}$~s$^{-1}$ for $E>10$~GeV and photon spectral index $\Gamma = 2.5$ as a function of sky location (in Galactic coordinates) for the 3FHL time interval. 
    \label{fig:sensitivity}}
\end{figure*}

\begin{figure}[!ht]
    \centering
    \includegraphics[width=8.2cm]{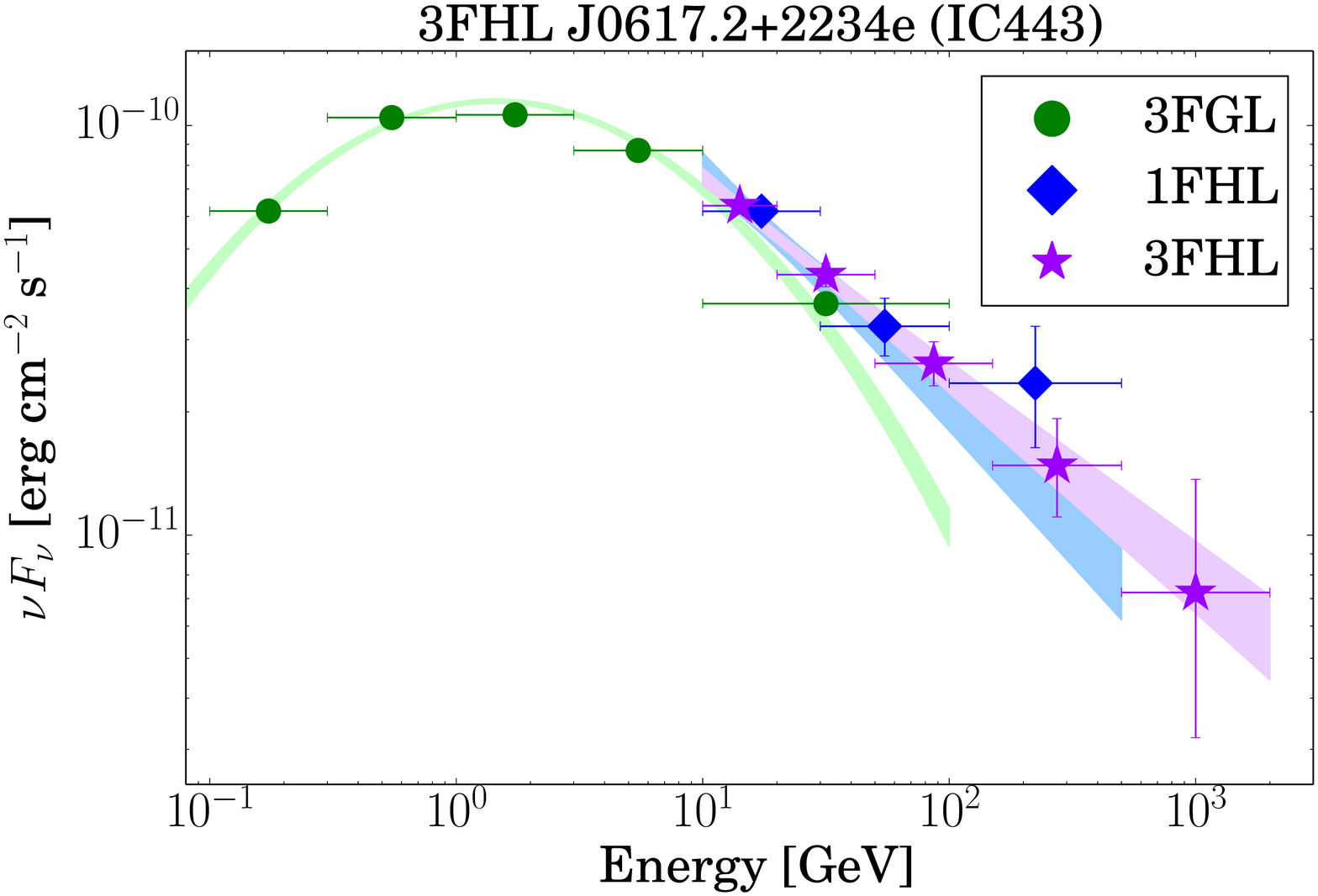}
    \includegraphics[width=8.2cm]{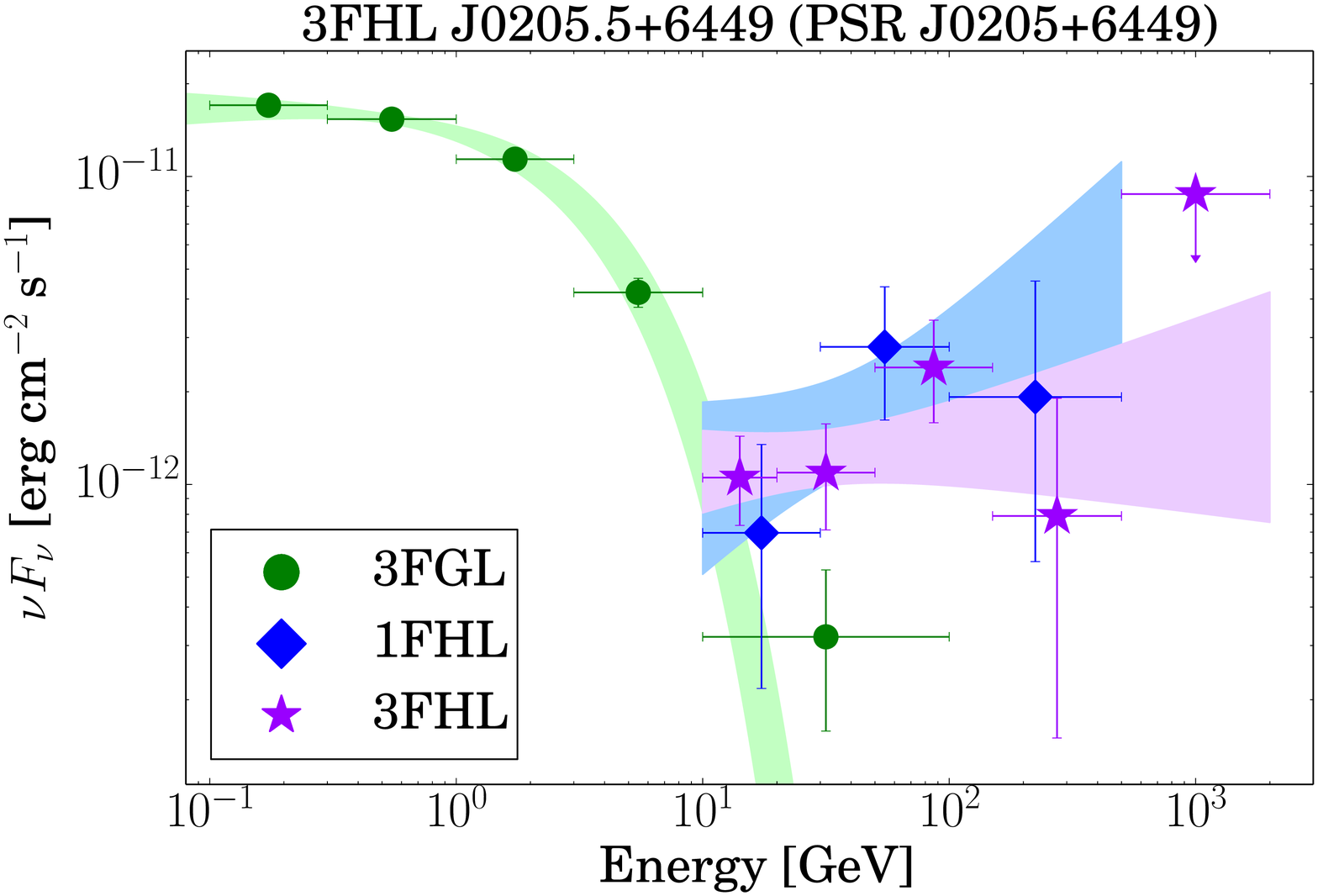}\\
	\includegraphics[width=8.2cm]{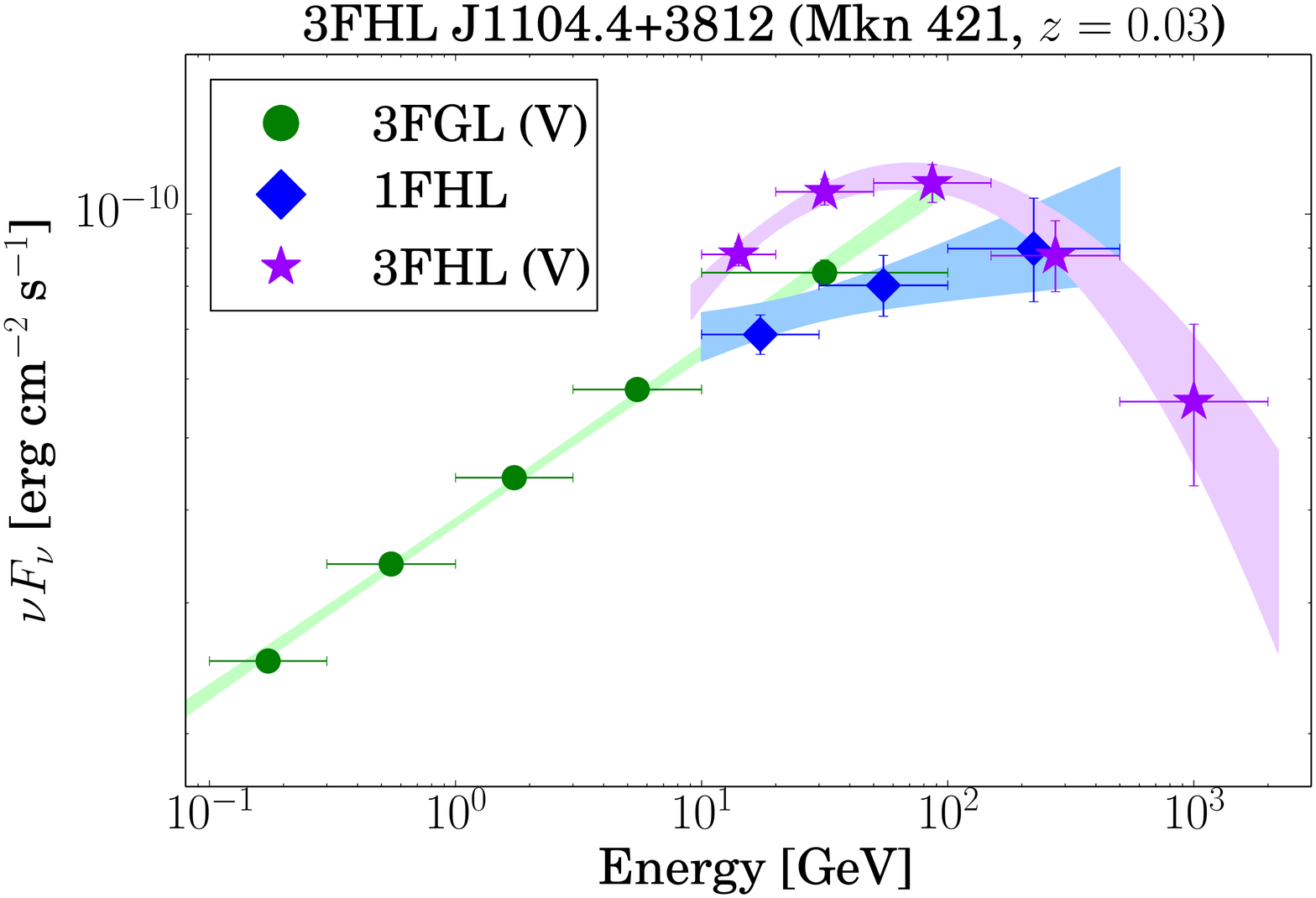} 
    \includegraphics[width=8.2cm]{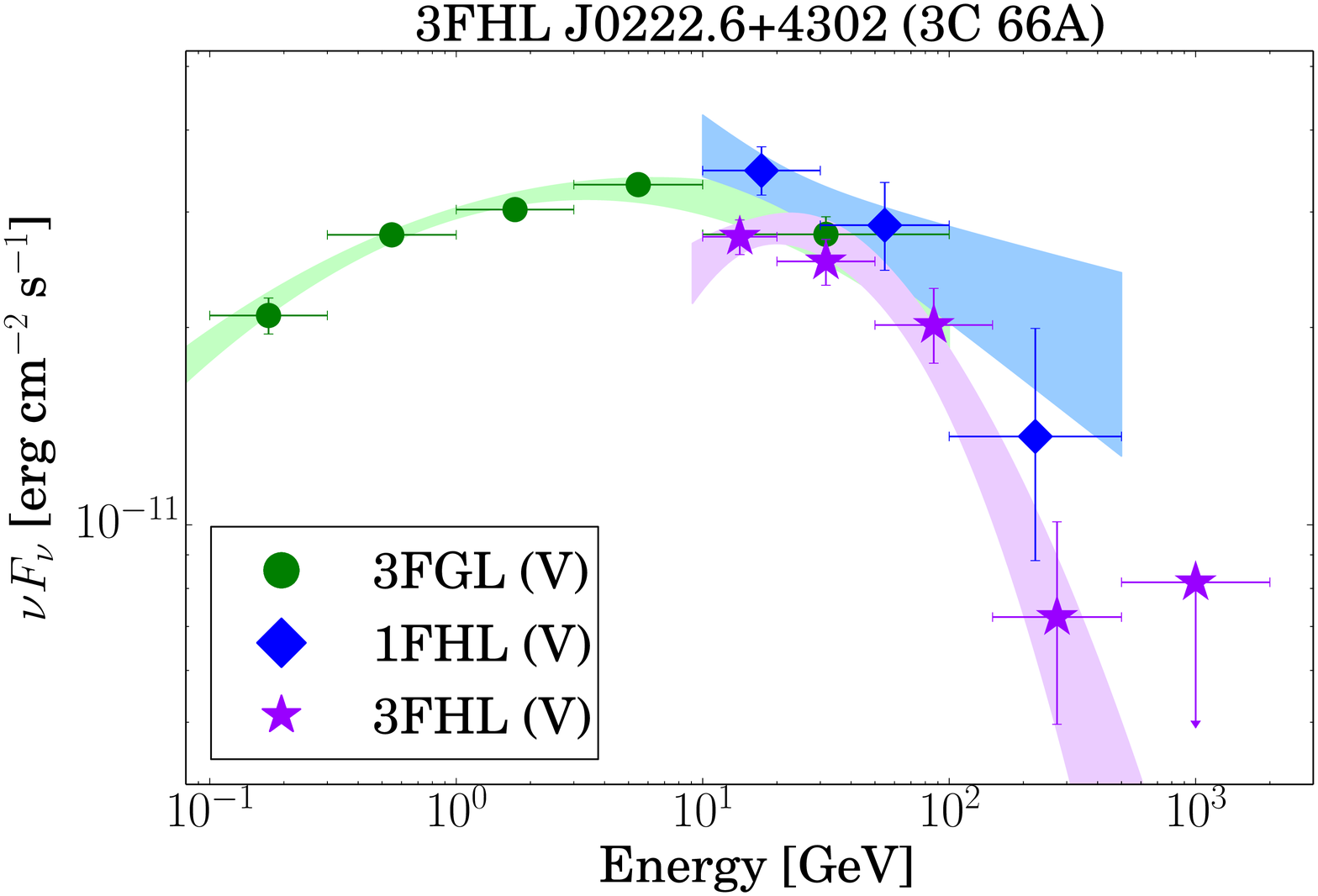}  
    \caption{Examples of SEDs for 3FHL sources. We combined the spectral data from the 3FGL (green circles) and 3FHL (magenta stars) to provide spectral coverage over four orders of magnitude. The 1FHL data (blue diamonds) are shown for comparison when available. The (V) stands for variable source according to the criteria in the respective catalog. We note that the SEDs of Mkn~421 (lower left panel) and 3C~66A (lower right panel) are characterized by a log-parabola shape. In these cases, a curved model is preferred over a power law at a significance of $3.1\sigma$ and $3.3\sigma$, respectively.
    \label{fig:seds}}
\end{figure}

%%%%%%%%%%%%%%%%%%%%%%%%%%%%%%%%%%%%%%%%%%%%%%%%%%%%%%%%%%%%%%%%
%
%         1FHL comparison
%
%%%%%%%%%%%%%%%%%%%%%%%%%%%%%%%%%%%%%%%%%%%%%%%%%%%%%%%%%%%%%%%%

\subsection{Comparison with the 1FHL Catalog}
In this section, we compare the 3FHL results with those of the previous \lat catalog at similar energies \citep[\ie the 1FHL,][]{1FHL}.

\begin{figure*}[!ht]
    \centering
    \includegraphics[width=8cm,trim=0 0 0 0cm]{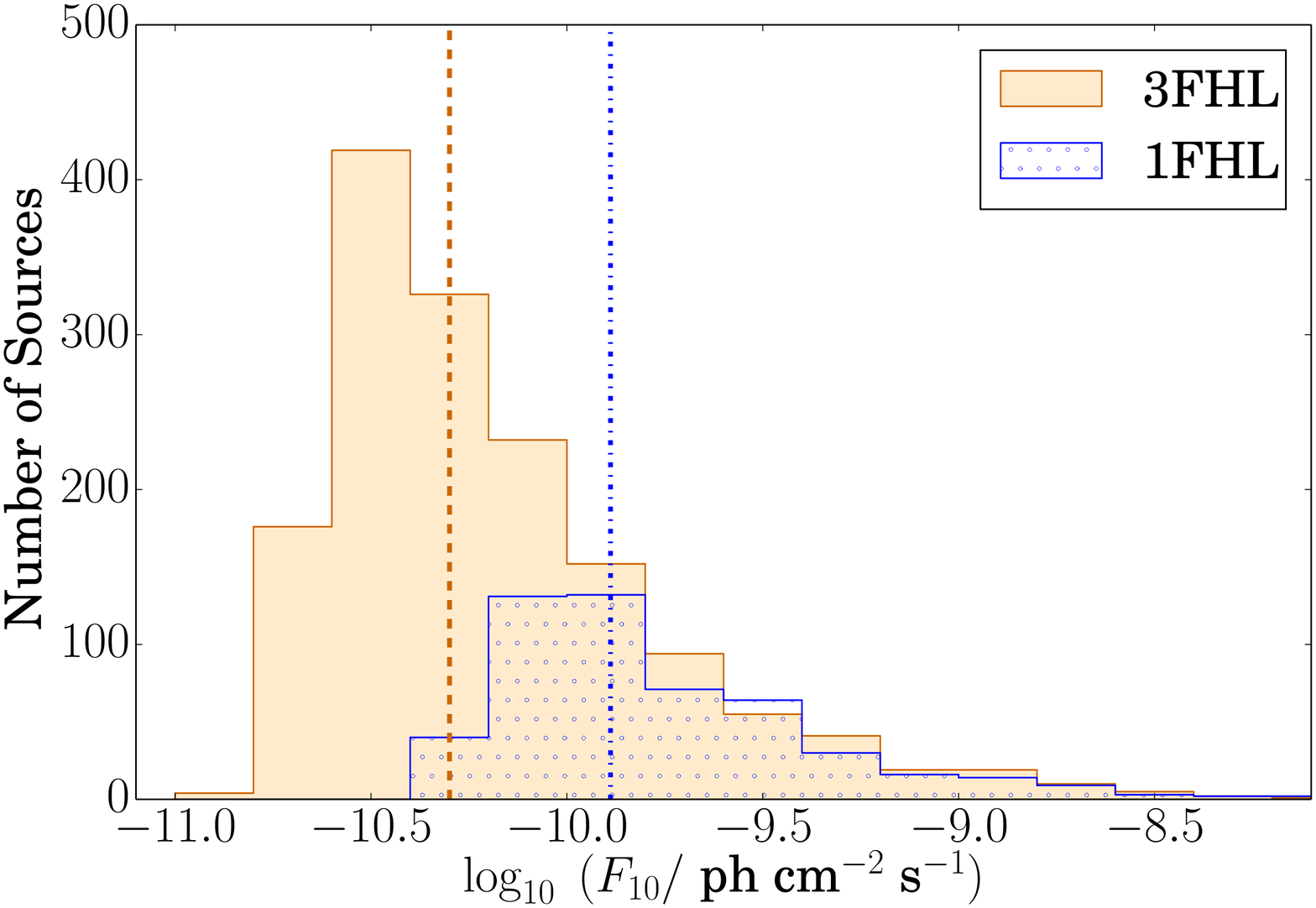} 
    \includegraphics[width=8cm,trim=0 0 0 0cm]{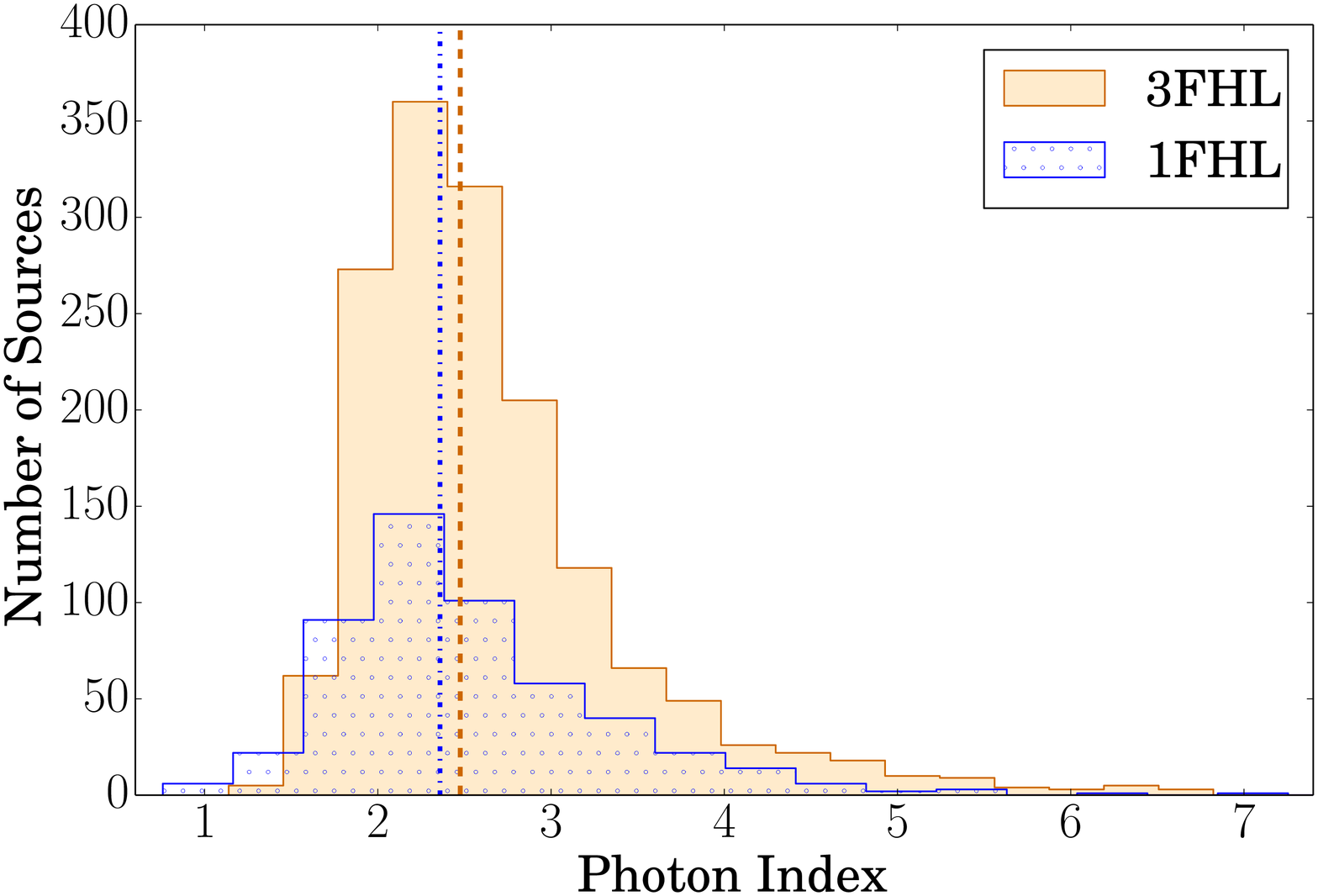}\\
    \includegraphics[width=8cm,trim=0 0 0 0cm]{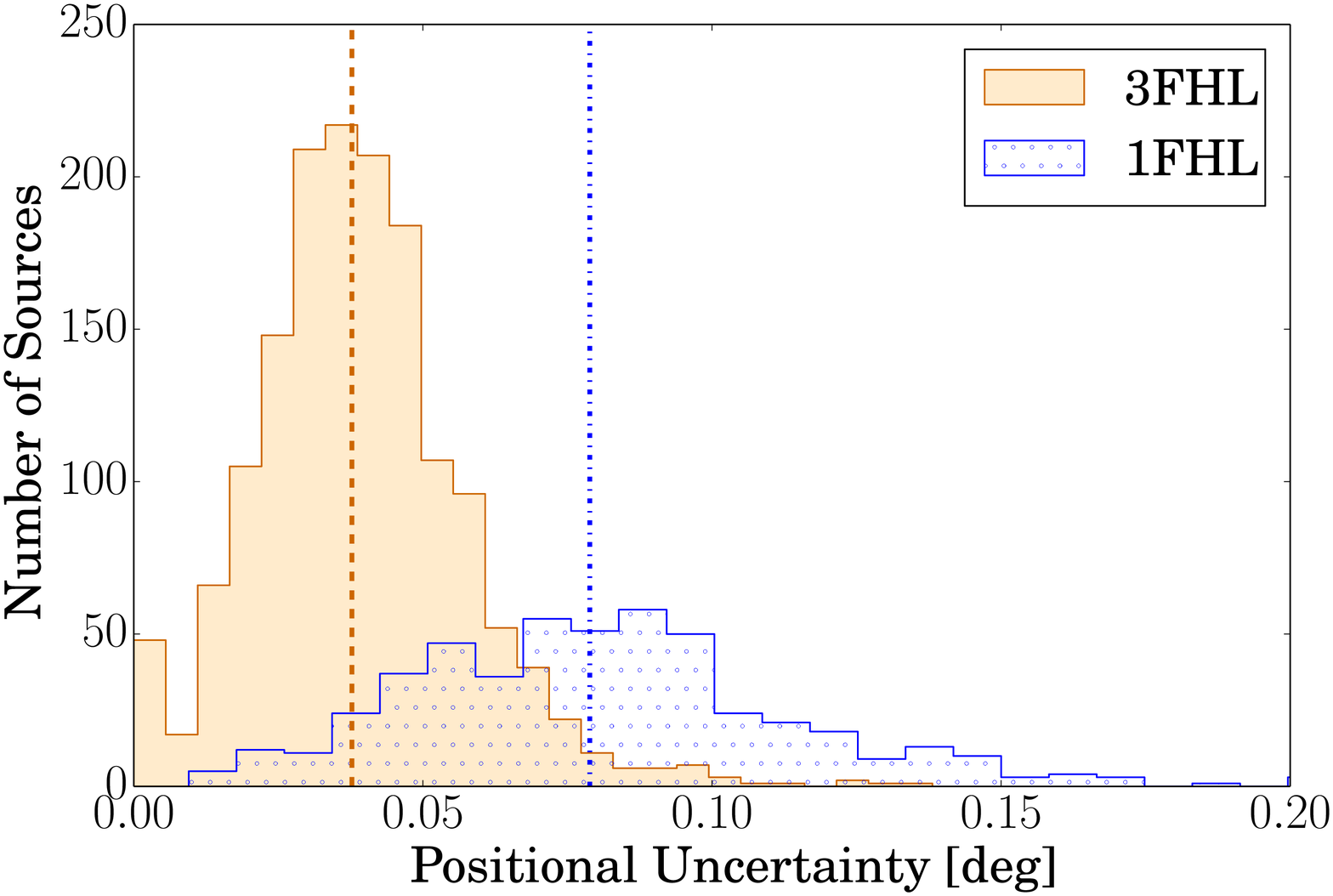} 
    \includegraphics[width=8cm,trim=0 0 0 0cm]{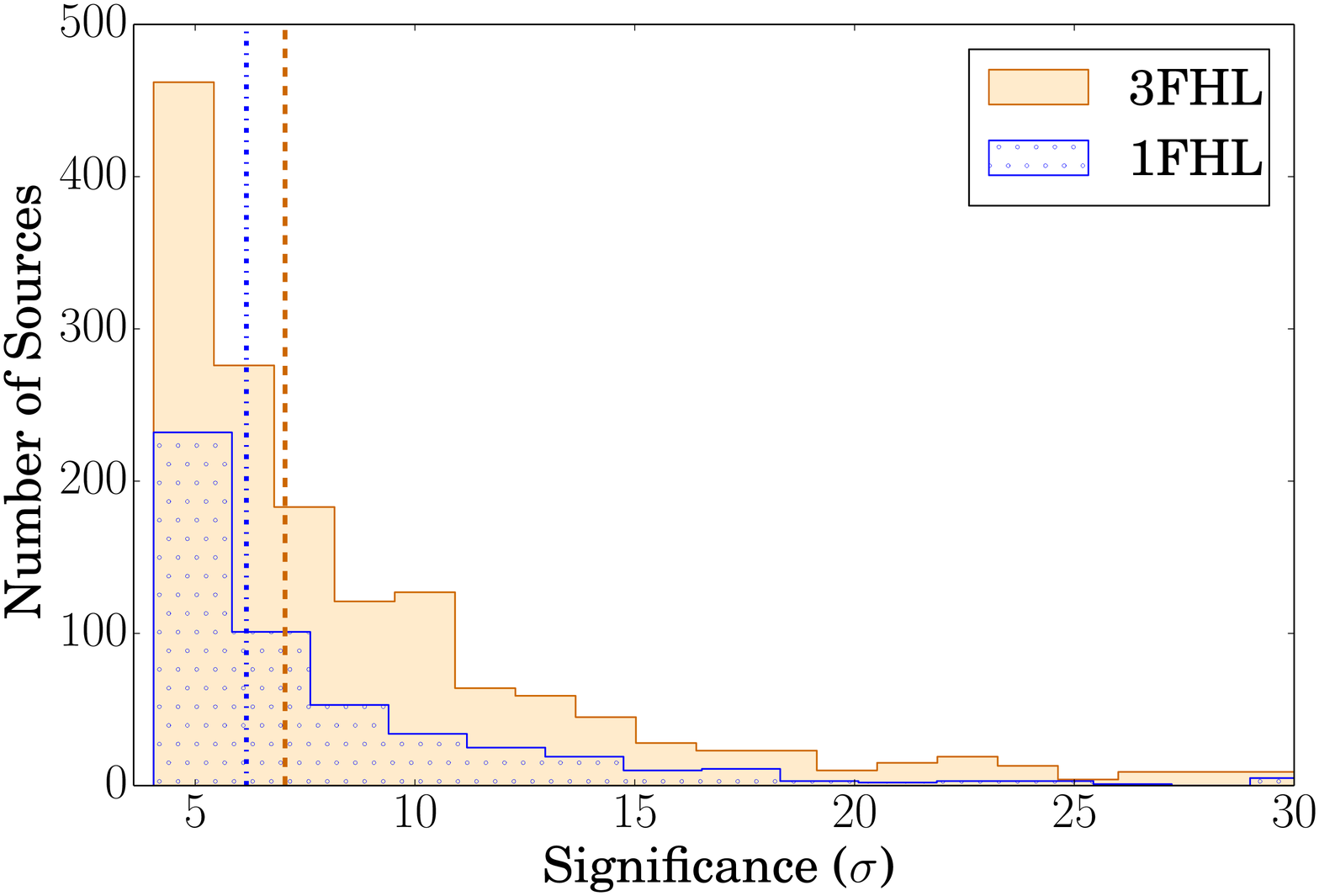} 
    \caption{Distributions of properties of 3FHL and 1FHL sources. ({\it Upper Left panel}) integrated flux above 10~GeV, ({\it Upper Right panel}) spectral index, ({\it Lower Left panel}) positional uncertainty (95\% error radius) and ({\it Lower Right panel}) significance of detection. The medians are shown with vertical lines.
    \label{fig:compa1fhl}}
\end{figure*}

The 1FHL was based on the first 3 years of data and the Pass~7 event reconstruction and classification analysis. For 3FHL we have analyzed 7 years of data using Pass~8. The total number of detected sources has increased by a factor of three, from 514 to 1556. A simple scaling of the sensitivity, assuming a background-limited scenario, would suggest $\leq 1000$ sources in 3FHL. The much larger number of sources detected in 3FHL shows that the sensitivity, for $\geq$10\,GeV, improves nearly linearly with time\footnote{We have assumed that the fluxes of the sources are distributed as a Euclidean $\log$ N-$\log$ S, \ie $N(>F)$ $\propto F^{-3/2}$, where $N$ is the number of sources above a given flux $F$.}. This is because the \lat operates in a counts-limited regime at these energies. This is demonstrated by the comparison of the flux distributions of the 1FHL and 3FHL sources in Figure~\ref{fig:compa1fhl}. Indeed a decrease in the median flux of a factor about 3, from $1.3\times 10^{-10}$ to $5.0\times 10^{-11}$~ph~cm$^{-2}$~s$^{-1}$ is apparent. The median of the spectral index distributions remains similar in both catalogs. In Figure~\ref{fig:compa1fhl}, we see that the 3FHL increases the size of the population of hard sources ($\Gamma\sim 1.8$) that was discovered in 1FHL. These are faint and hard HSP BL~Lacs (see \S~\ref{sec:ext} for more details) that are detected in 3FHL because of the improved sensitivity at high energies delivered by Pass~8. Furthermore, Figure~\ref{fig:compa1fhl} also compares the distributions of positional uncertainties. There is a clear improvement in the median positional resolution by approximately a factor of 2, from 4.7 to 2.3~arcmin, 95\% C.L. This is better than the 3.8 arcmin median location uncertainty (95\% C.L.) of 2FHL \citep{2FHL} thanks to the generally larger statistics existing at $>10$\,GeV for 3FHL sources. Figure~\ref{fig:compa1fhl} also shows the distributions of detection significance for both catalogs.

Sixteen 1FHL sources are missing in 3FHL. Among these only one was very significant, 1FHL~J1758.3$-$2340 with $\mathrm{TS}=47$.  In 3FHL this source is now part of the FGES J1800.5$-$2343 disk, which is broader than W~28 was in 1FHL. The rest of the missing sources had a TS between 25 and 30. Three of the missing 1FHL sources are parts of new extended sources: 1FHL~J0425.4+5601 and 1FHL~J0432.2+5555 in FGES J0427.2+5533 and 1FHL~J1643.7$-$4705 in FGES J1633.0$-$4746. Only one other missing 1FHL source (1FHL J1830.6$-$0147) had $N_{pred}>4$. It coincides with a cluster of four 3FGL sources, so it is possibly an extended source remaining to be discovered. All of these sources have corresponding seeds in the 3FHL analysis pipeline that were rejected from the catalog for having a TS~$\sim 15$. We stress that the 1FHL catalog was built from Pass~7 Clean class events (before reprocessing). Therefore the set of events was rather different, and the PSF was significantly broader.

%%%%%%%%%%%%%%%%%%%%%%%%%%%%%%%%%%%%%%%%%%%%%%%%%%%%%%%%%%%%%%%%
%
%         New sources
%
%%%%%%%%%%%%%%%%%%%%%%%%%%%%%%%%%%%%%%%%%%%%%%%%%%%%%%%%%%%%%%%%

\subsection{New $\gamma$-ray Sources and TeV Candidates}
\label{newandTeV}

\begin{figure*}[!ht]
    \centering
    \includegraphics[width=16cm,trim=0 0 0 0cm]{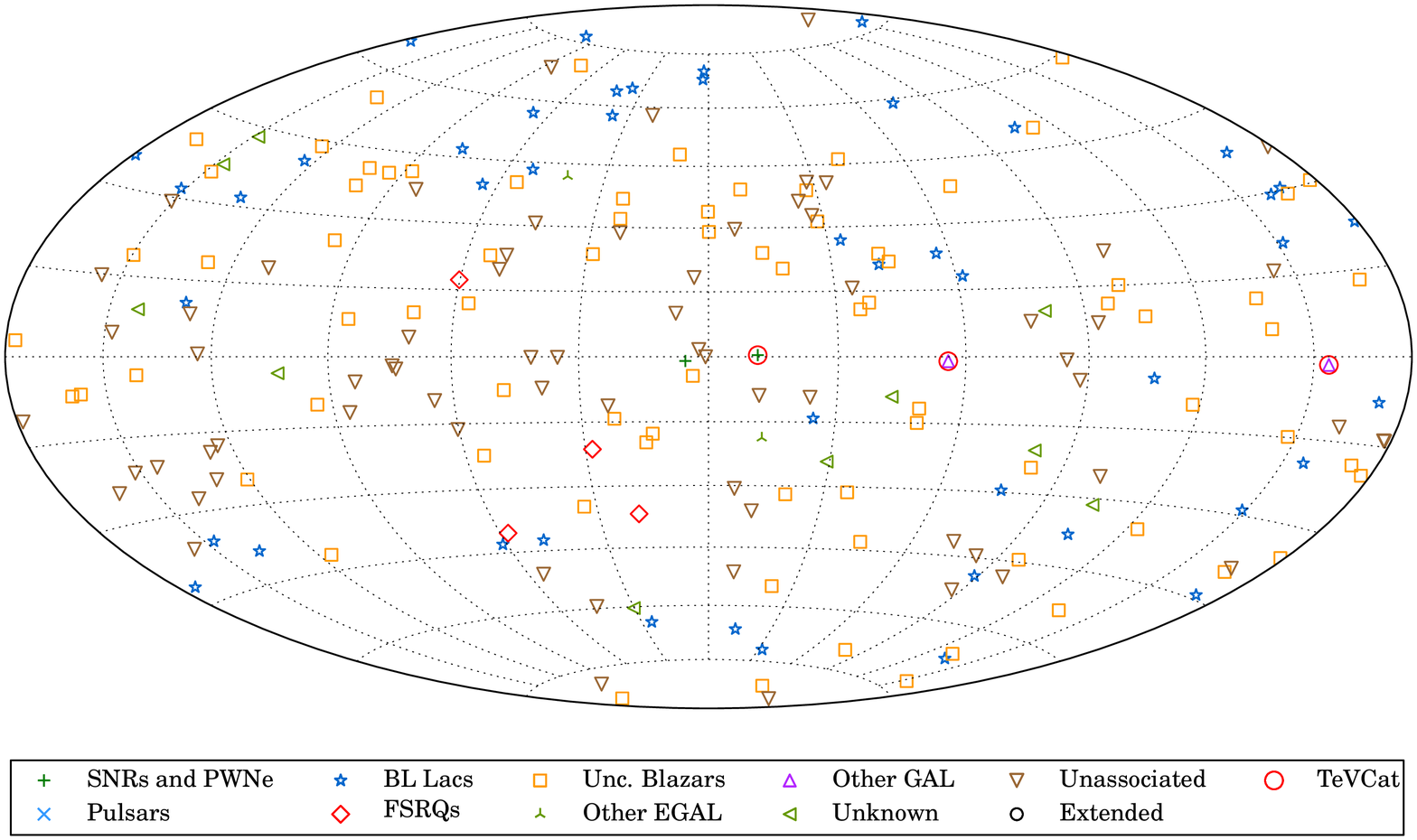} 
    \caption{Sky map, in Galactic coordinates and Hammer-Aitoff projection, showing the 214 objects in the 3FHL catalog previously undetected by the LAT. The sources are classified by their most likely source class. The 3 sources already found by IACTs are indicated with open red circles. 
    \label{fig:all_sky-new}}
\end{figure*}

Thanks to the unprecedented low flux limit of our analysis at $E>10$~GeV, the 3FHL analysis has revealed a large number of new sources. The number of 3FHL sources without a 3FGL counterpart is 258, and of these 214 have no counterparts in any previous \lat catalog. Three of these 214 have been detected with IACTs, \ie 3FHL J0632.7+0550 \citep[HESS J0632+057, ][]{caliandro15}, 3FHL J1303.0$-$6350 (PSR~B1259$-$63, which flared after the 3FGL time period), and 3FHL J1714.0$-$3811 (CTB~37B, previously unresolved). In summary, the 3FHL has 211 sources previously unknown in $\gamma$ rays. The sky locations and classes of the sources not previously detected by the LAT are shown in Figure~\ref{fig:all_sky-new}. This figure shows that while most of them appear to be isotropically distributed, some fraction of the unassociated new sources appears to lie in the Galactic plane (see \S\ref{sec:galactic}).

\begin{figure}[!ht]
    \centering
    \includegraphics[width=\columnwidth]{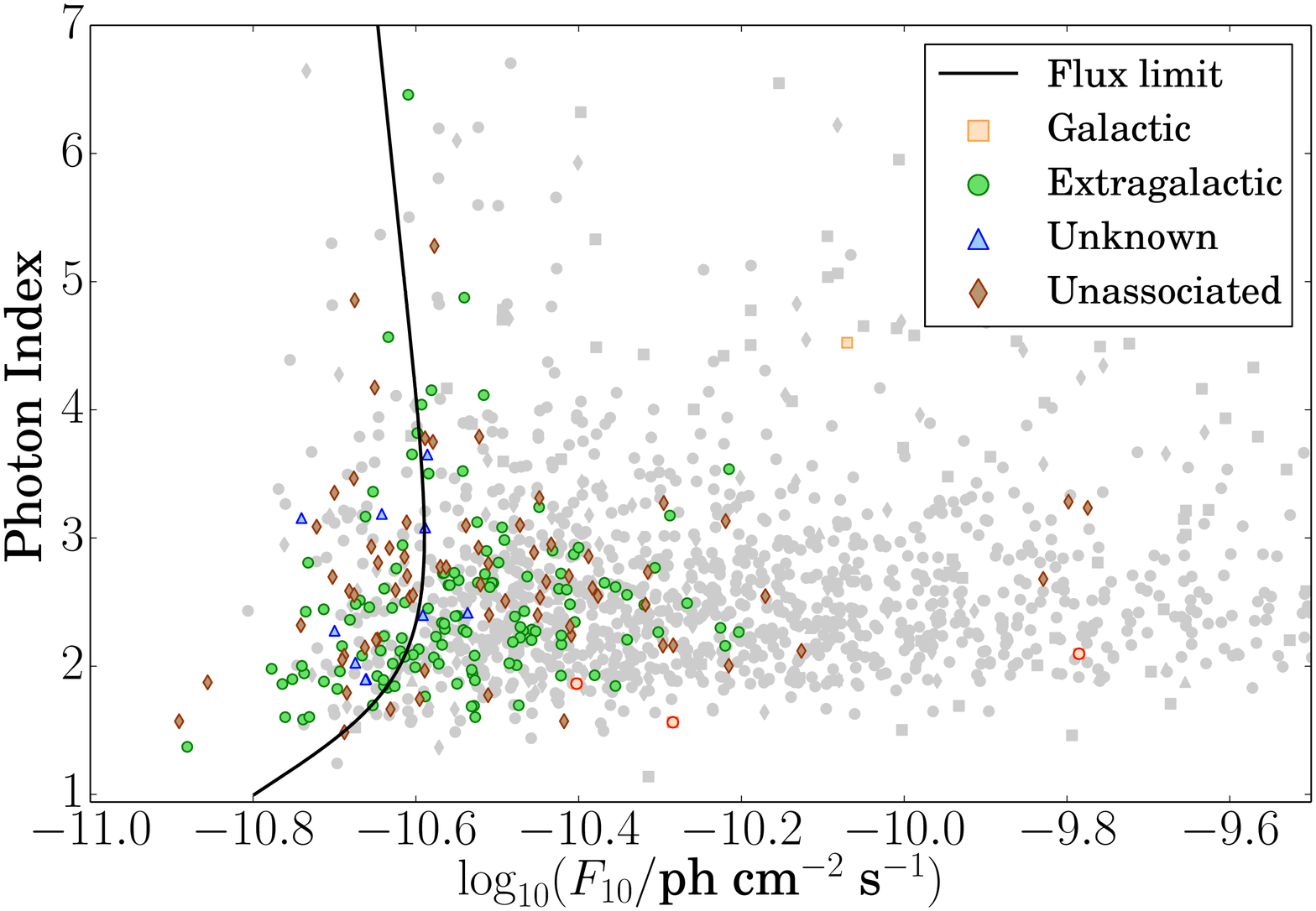} 
    \caption{The spectral index of Galactic (orange squares), extragalactic (green circles), unknown (blue triangles) and unassociated sources (brown diamonds) versus the integrated flux above 10~GeV for the 3FHL sources not in previous \lat catalogs. The black line shows the flux limit averaged over the high latitude sky ($|b|\geq 10^{\circ}$). Symbols with a red outline are sources already detected at TeV energies and contained in the TeVCat catalog. For comparison, the rest of the 3FHL sources are shown as gray symbols.
    \label{fig:index_vs_fluxs}}
\end{figure}

Figure~\ref{fig:index_vs_fluxs} shows that the new sources follow an index distribution similar to that of the previously detected sources but are fainter. Most new sources are either extragalactic or unassociated (but probably extragalactic). The new Galactic sources tend to be already known to be spatially extended from IACT observations, and we model them as such.

%IACTs are characterized by an excellent flux sensitivity at TeV energies but a limited FoV that makes challenging for them to find new sources. The 3FHL provides a great resource to plan Cherenkov observations. The catalog lists the highest-energy photon (HEP) detected by the LAT that is coming from a given source. Sources with HEPs of $\sim 100$s of GeV, small indexes (hard spectra), and bright are, a priori, good candidates for IACTs.

IACTs have excellent flux sensitivities at TeV energies but limited fields of view that make finding new sources challenging. The 3FHL is a resource for planning IACT observations. The catalog lists the highest-energy photon (HEP) detected by the LAT and its probability of association with a given source. Sources with HEPs of hundreds of GeV, small indices (hard spectra), and large fluxes are, a priori, good candidates for IACTs. However, the majority of the $\gamma$-ray sources detected by the LAT in 3FHL may be too faint for the current-generation IACTs, which can reach a sensitivity of 2.7$\times 10^{-12}$~erg~cm$^{-2}$~s$^{-1}$ (this is, $>1-2$\,\% of the Crab Nebula flux at $\gtrsim 100$\,GeV in 50 hours of observation). Of the 1423 3FHL sources that have not been detected by IACTs, only 8 have $>$100\,GeV flux $>10^{-12}$\,erg~cm$^{-2}$~s$^{-1}$ ($\sim$0.3\,\% of the Crab Nebula flux). Of those, 6 are already reported in 2FHL, while the remaining two are extended sources (3FHL~J1036.3$-$5833e and 3FHL~J1824.5$-$1351e) in the Galactic plane. These two sources are the brightest among the above group with $>$100\,GeV flux of $\sim30$\,\% and $\sim50$\,\% of the Crab Nebula flux, respectively and HEPs of $\sim$355\,GeV and 586\,GeV. Thus, hard Galactic sources, with limited extension, may be the best targets for current-generation IACTs.

In the \texttt{TEVCAT\_FLAG} column of the catalog, we have flagged the sources considered as good TeV candidates based on these criteria \citep[which are from ][see \S5.1 of that paper]{1FHL}: (1) the source significance above 50~GeV is $\sigma_{50}>3$, (2) the power-law spectral index above 10~GeV is $\Gamma<3$, and (3) the integrated flux above 50~GeV is $F_{50}>10^{-11}$~ph~cm$^{-2}$~s$^{-1}$. This selection results in 246 candidates for TeV detection.

%% Marco: I computed the Crab >100 GeV flux from the MAGIC paper as:
% TF1 *a=new TF1("crab","3.39e-11*TMath::Power(x,-2.51-0.21*log10(x))*x",0.065,100)
% a->Integral(0.1,100)*1.6 = 2.7e-10 erg/cm2/s
% 2% Crab at >100 MeV is 5.4e-12 erg/cm2/
%%%%%%%%%%%%%%%%%%%%%%%%%%%%%%%%%%%%%%%%%%%%%%%%%%%%%%%%%%%%%%%%
%
%         Galactic Science
%
%%%%%%%%%%%%%%%%%%%%%%%%%%%%%%%%%%%%%%%%%%%%%%%%%%%%%%%%%%%%%%%%

\subsection{The Galactic Population}
\label{sec:galactic}
The majority of Galactic sources detected in 3FHL are sources at the final stage of stellar evolution such as pulsars, PWNe and SNRs, many of which are detected as extended, and high-mass binaries.

\begin{figure}[!ht]
    \centering
    \includegraphics[width=\columnwidth]{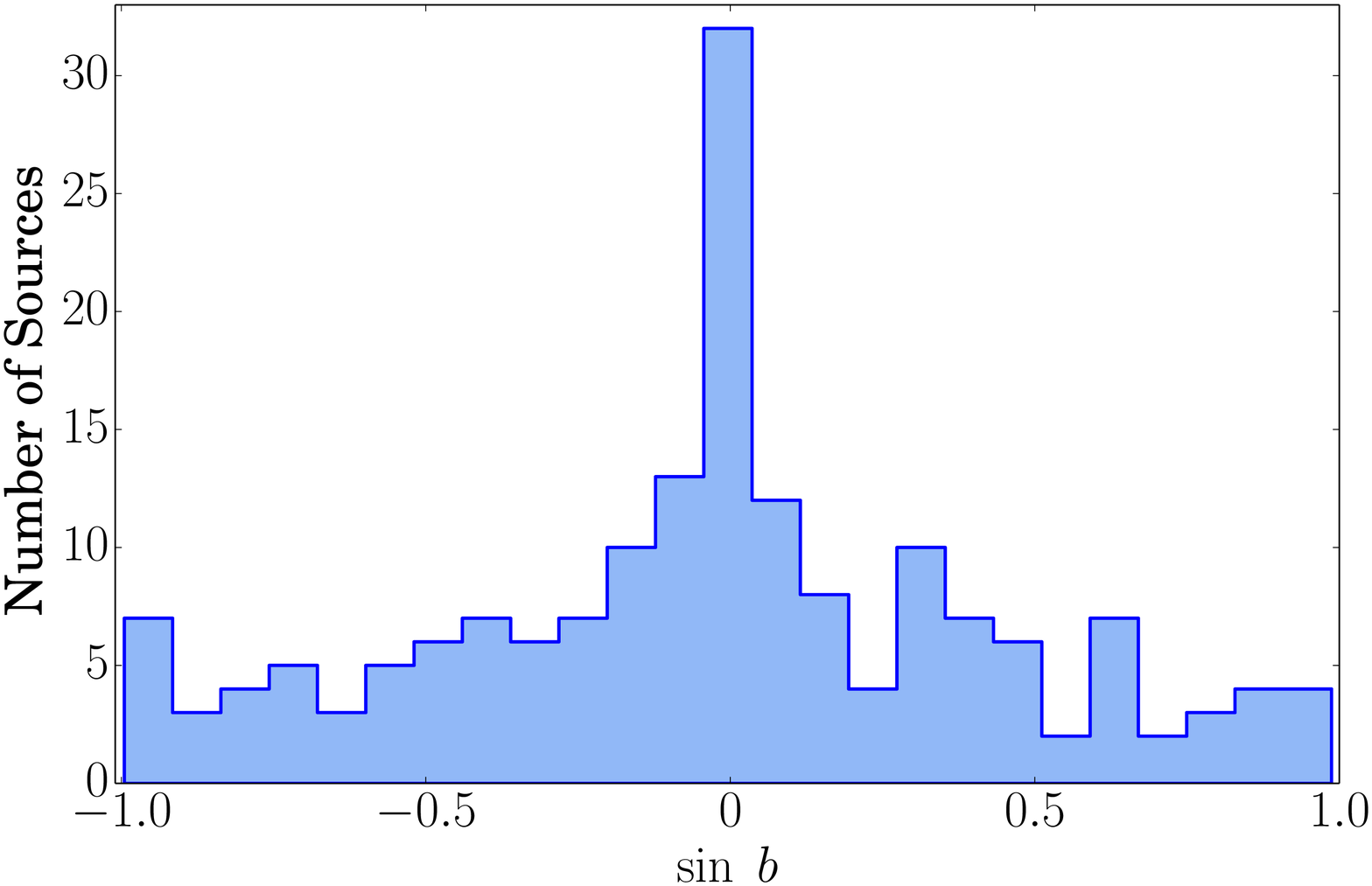} 
    \caption{Distribution of unassociated sources over the sine of the Galactic latitude. This distribution peaks at $|b|<2^{\circ}$ on top of an isotropic background of sources.    
    \label{fig:hist_sinb}}
\end{figure}

In this catalog 125 sources are associated with Galactic objects and 83 are unassociated within the plane of our Galaxy ($|b|<10^{\circ}$). The same low Galactic latitude region has 133 extragalactic objects. Considering the density of extragalactic sources outside of the plane and the decreased sensitivity for source detection in the plane we estimate that $\approx$25--40 of the 83 unassociated objects may be Galactic. Indeed, the distribution in Galactic latitude of unassociated sources (see Figure~\ref{fig:hist_sinb}) shows a peaked profile for $|b|<2^{\circ}$ on top of a flat isotropic background.
%%%% Density of blazars in the plane
% I used the number outsude and corrected for the sensitivy factor of 2 (ratio of the two median energy flux)
% 41253-30000.)/30000.*1387*pow(2,-1.5) =184 sources

\begin{figure}[!ht]
    \centering
    \includegraphics[width=\columnwidth]{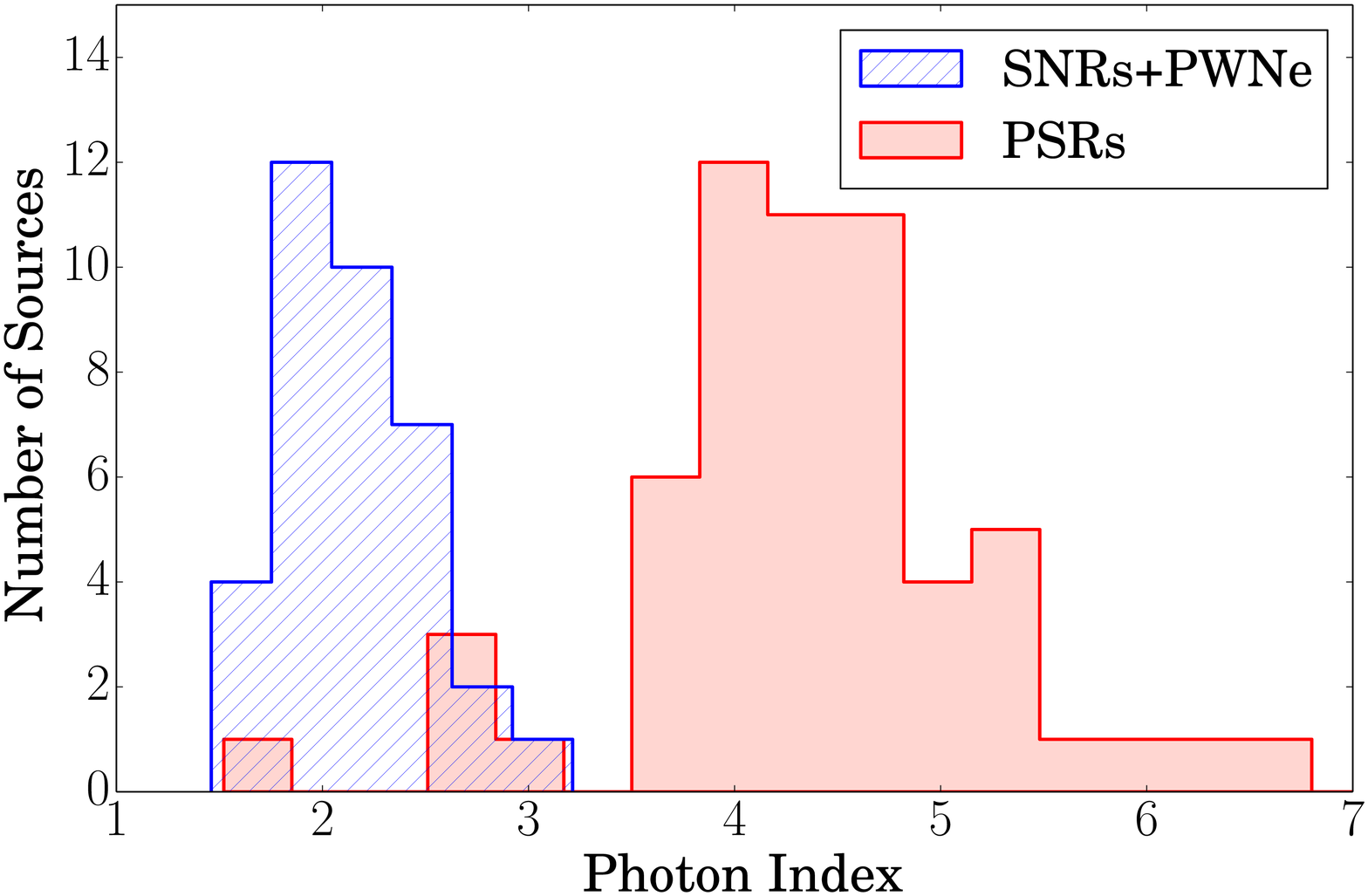} 
    \caption{Distributions of $\gamma$-ray spectral indices of SNRs plus PWNe (dashed blue) and sources associated with PSRs (filled red). At the 3FHL energies, SNRs and PWNe tend to have smaller indices (harder spectra) than PSRs, for which the LAT measurement is sensitive to the exponential cutoff.
    \label{fig:hist_galtype}}
\end{figure}

The spectral index distribution of Galactic sources is broad with a median index $\Gamma\approx 3$ as shown by Figure~\ref{fig:hist_galtype}. This arises from the superposition of the distributions of the indices of the different source classes. The majority of sources are pulsars and, at $>$10\,GeV, the LAT samples their super-exponential cutoffs yielding a median spectral index of $\Gamma\approx 4$. Sources classified as pulsars in 3FGL retain this classification in 3FHL for consistency. A source is reclassified as PWN only if it is associated with a known, small-size PWN and has a rising SED indicative of a dominant PWN component. Only 3FHL J0205.5+6449, 3FHL J0534.5+2201 and 3FHL J1124.4-5916 have been reclassified accordingly. SNRs and PWNe account for 56 objects. Their similar index distributions translate into much harder spectra than the rest, having a median of $\Gamma\approx 2$. The unassociated sources within the plane of the Galaxy display the full range of spectral indices $1< \Gamma < 5$. However, those within $|b|<2^{\circ}$ primarily have $\Gamma<2.5$ suggesting PWN or SNR natures. At latitudes $|b|\geq 10^{\circ}$ the 3FHL catalog contains 15 millisecond pulsars (MSPs) of which 13 are classified as PSR (discovered pulsating in $\gamma$ rays by the LAT) and 2 as psr (radio MSP with no detection of $\gamma$ pulsations). We also find three young pulsars classified as LAT PSR and one PWN in the LMC (N 157B).

\begin{figure*}[!ht]
\begin{center}
\begin{tabular}{ll}
        % original scale=0.4
        \includegraphics[angle=90,scale=.3]{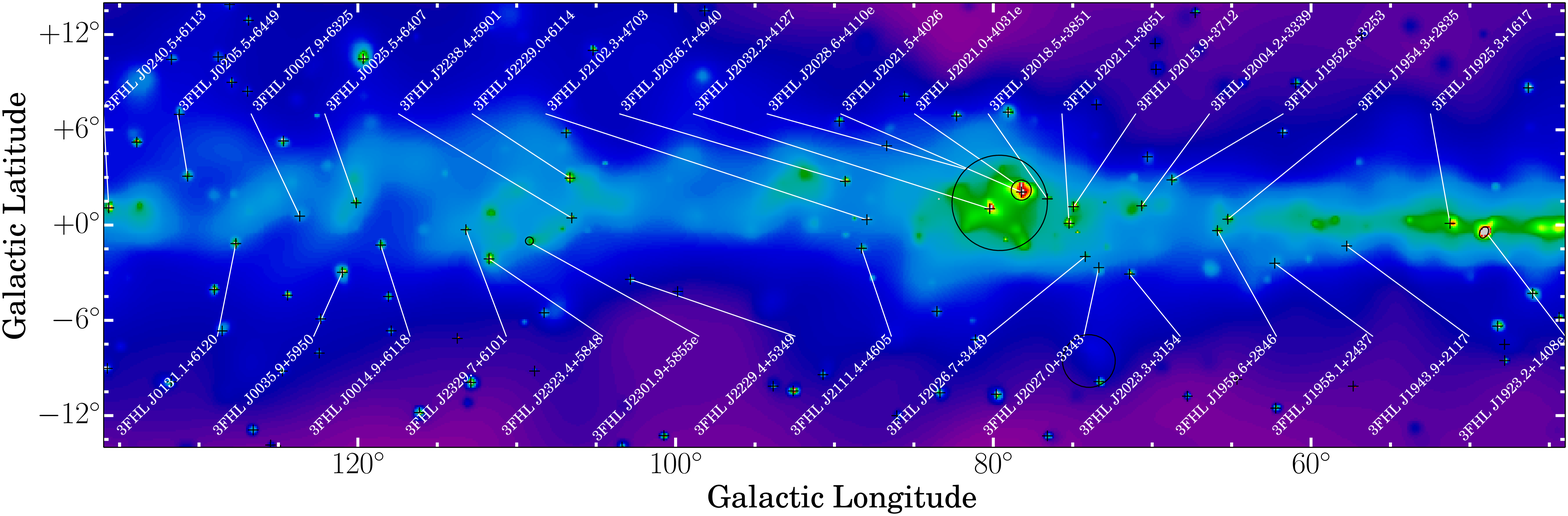}&
        \includegraphics[angle=90,scale=.3]{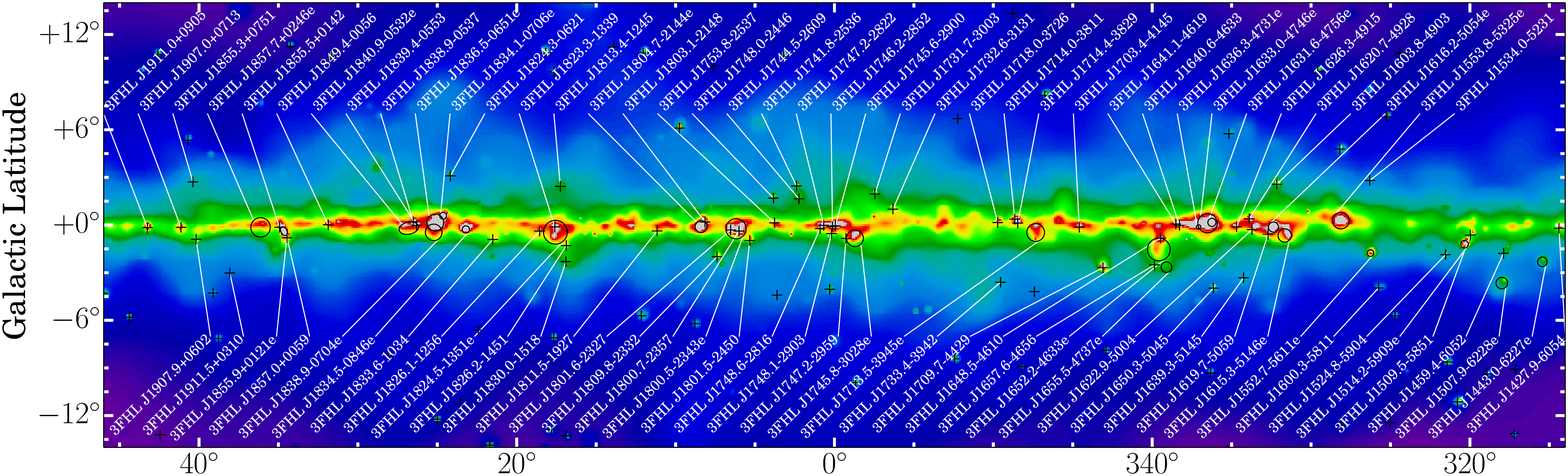}\\
\end{tabular}
\end{center}
    \caption{Adaptively smoothed counts map showing the whole Galactic plane $0^{\circ}\leq l\leq 360^{\circ}$ at Galactic latitudes $-14^{\circ} \leq b\leq 14^{\circ}$ divided in four  panels. The panels are centered at $l=90^{\circ}$, $0^{\circ}$, $270^{\circ}$ and $180^{\circ}$, from top to bottom. Detected point sources are marked with a cross whereas extended sources are indicated with their extensions. Only sources located at $-4^{\circ} \leq b\leq 4^{\circ}$ are labelled, plus the Crab Nebula.
    \label{fig:gp1}}
\end{figure*}

With respect to the 1FHL catalog, the 3FHL doubled the number of Galactic objects detected at $>$10\,GeV, maintaining similar proportions among the source classes. On the other hand the Galactic sources in the 2FHL catalog, because of its $>50$ GeV threshold, are primarily PWNe/SNRs, with only one pulsar. Within the region $|b|\leq5^{\circ}$, which is where the diffuse flux is the brightest, the sensitivity of the 3FHL analysis reaches a median of $\sim5\times10^{-12}$\,erg cm$^{-2}$ s$^{-1}$; this is $\sim$1\,\% of the Crab Nebula flux in the 10\,GeV -- 2\,TeV band. Transformed to the energy range $>$1\,TeV based on the Crab Nebula spectrum, this corresponds to an energy flux of $\sim 8\times10^{-13}$\,erg cm$^{-2}$ s$^{-1}$, which is slightly better than the sensitivity of $\sim 1.4\times 10^{-12}$\,erg cm$^{-2}$ s$^{-1}$ reached at $>$1\,TeV by H.E.S.S. in its Galactic plane survey \citep{aharonian06_gps,carrigan2013}. Within the footprint\footnote{The H.E.S.S. Galactic survey extends over the range $283^{\circ} < l < 59^{\circ}$ and Galactic latitudes of $|b| < 3\fdg5$.} of the H.E.S.S. survey, where H.E.S.S. detects 69 objects (reported in TeVCat\footnote{\url{http://tevcat.uchicago.edu/}}), the LAT detects 111 objects, of which 43 are in common with H.E.S.S. Detections at TeV energies are related to the spectral hardness. Indeed, the median spectral index of 3FHL sources detected in the H.E.S.S. survey is $\sim$2.0, while it is $\sim$2.5 for those that are undetected. Cut-out images of the Galactic plane are shown in Figure~\ref{fig:gp1}.

Of the 15 hardest sources ($\Gamma\leq 1.7$) detected at latitudes $|b|<10^{\circ}$, only four and seven are detected in TeVCat and 2FHL, respectively. There are five objects associated with Galactic classes, four blazars, and six unassociated. None of the blazars are in the TeVCat, maybe due to source activity. Variability cannot affect the comparison between the 3FHL and 2FHL because they span essentially the same time period. Indeed all of the 3FHL AGNs located in the Galactic Plane were included in 2FHL.

\begin{figure*}[!ht]
\begin{center}
\begin{tabular}{ll}
        \includegraphics[angle=90,scale=.3]{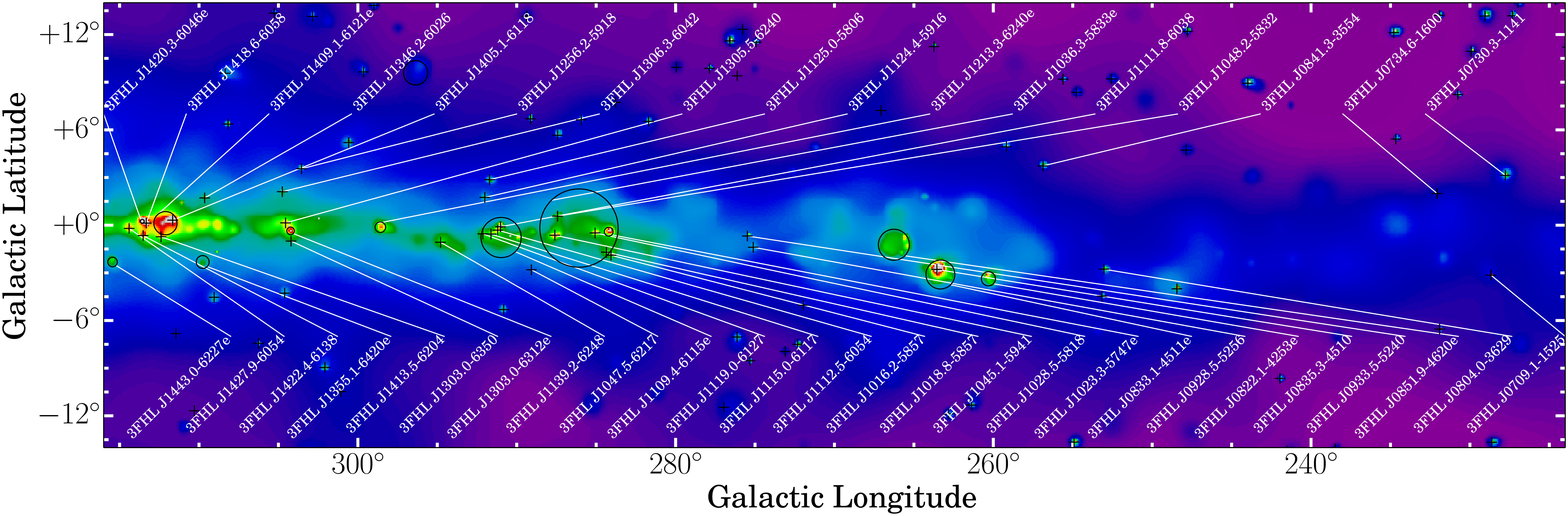}&
        \includegraphics[angle=90,scale=.3]{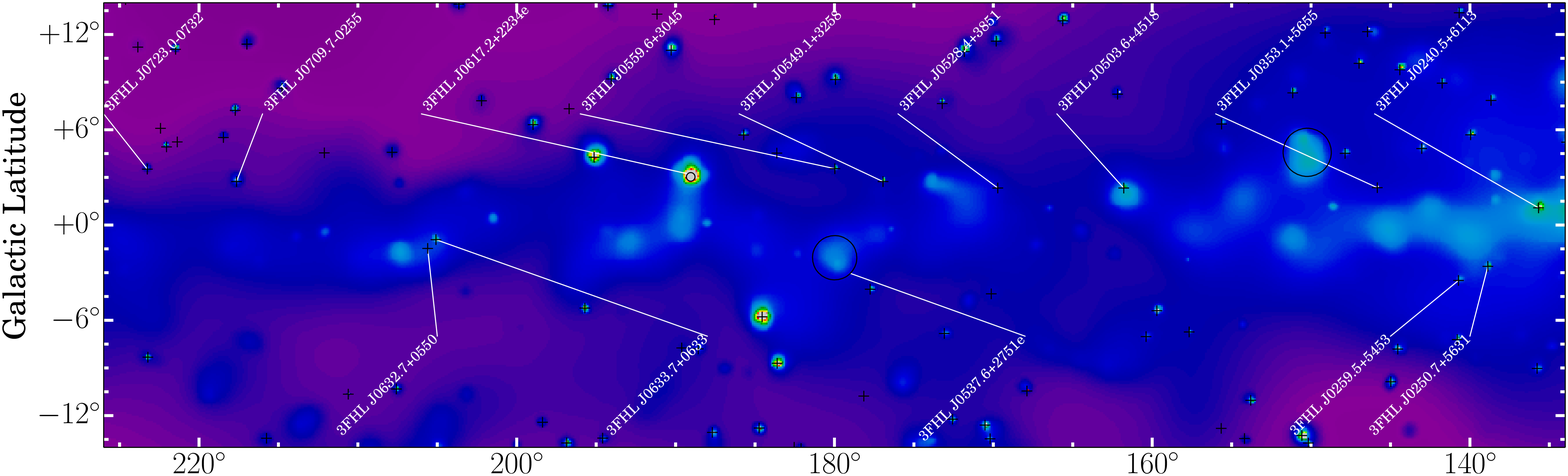}\\
\end{tabular}
\end{center}
    \begin{flushleft}
    {Fig.~\ref{fig:gp1}.}---  continued
    \end{flushleft}
\end{figure*}

%%%%%%%%%%%%%%%%%%%%%%%%%%%%%%%%%%%%%%%%%%%%%%%%%%%%%%%%%%%%%%%%
%
%         Extragalactic Science
%
%%%%%%%%%%%%%%%%%%%%%%%%%%%%%%%%%%%%%%%%%%%%%%%%%%%%%%%%%%%%%%%%
\subsection{The Extragalactic Population} \label{sec:ext}
The sky above 10~GeV is dominated by extragalactic sources (1231 sources, 79\% of the whole catalog). Blazars are the most numerous source type. We find associations with 750 BL Lacs, 172 FSRQs, 290 BCUs and 19 extragalactic sources with a different classification (representing 61\%, 14\%, 23\% and 2\% of the total extragalactic sky, respectively). These results differ from what was found at $>$50~GeV (\ie 2FHL). In the 2FHL, 65\% of the sources with $|b|>10^{\circ}$ were associated with BL Lacs (mostly HSP BL Lacs) and only 4\% with FSRQs. However, at $>$10~GeV there is a more diverse AGN population, confirming that a strong spectral cutoff in the range 10--50~GeV is common.

\begin{figure}[!ht]
    \centering
    \includegraphics[width=\columnwidth]{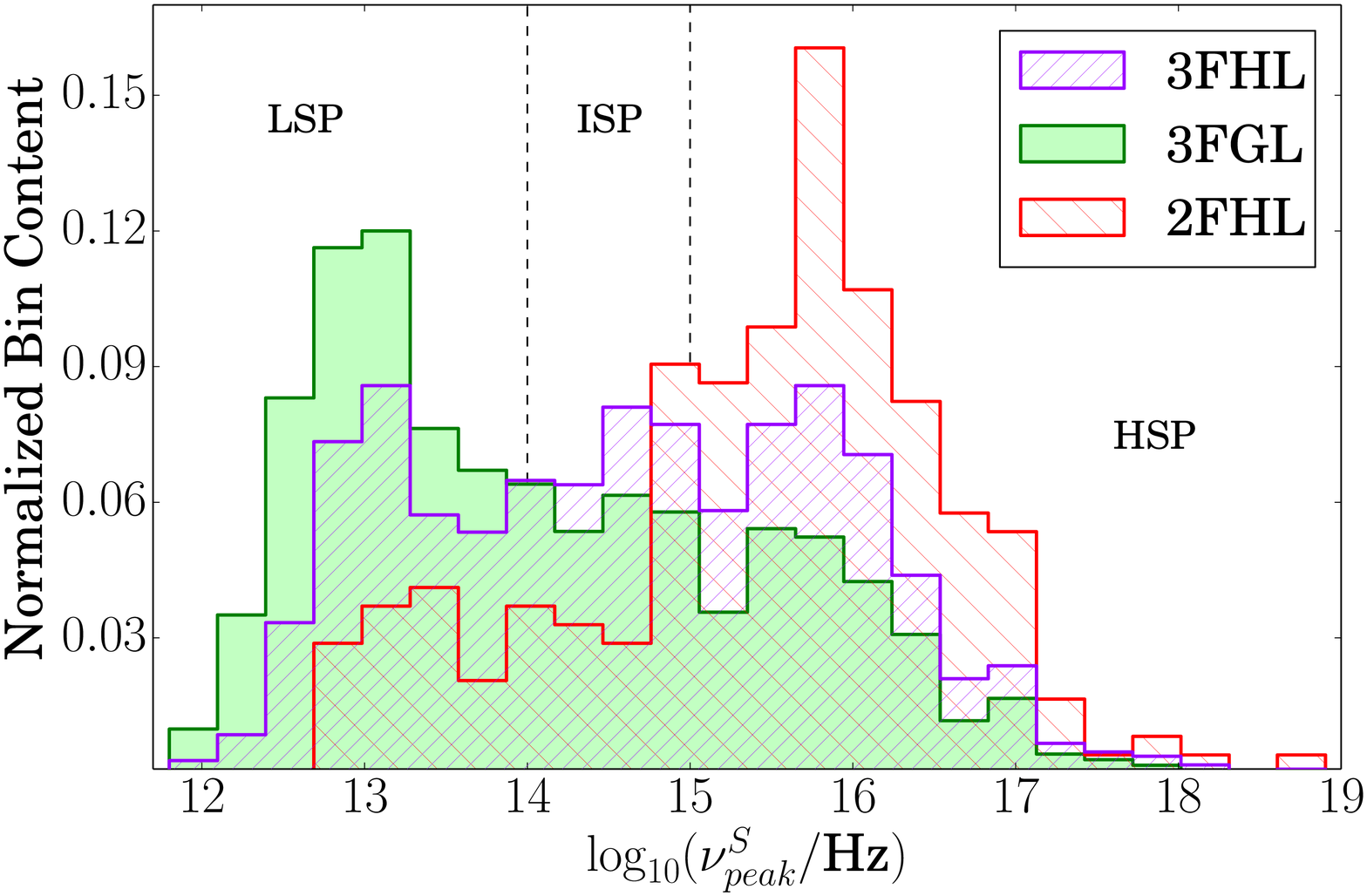} 
    \caption{Normalized distributions of the frequency of the synchrotron peak for the blazars detected in the 3FGL (0.1--300~GeV), 2FHL (50~GeV--2~TeV), and 3FHL (10~GeV--2~TeV) catalogs.
    \label{fig:nupeak}}
\end{figure}

Figure~\ref{fig:nupeak} shows the distribution of synchrotron peak frequencies of blazars detected in the 3FGL, 2FHL, and 3FHL \citep{sed_abdo10}. The 3FGL and 2FHL catalogs clearly sample different parts of the blazar population, with the 3FGL including mostly LSPs and ISPs and the 2FHL including mostly HSPs. The 3FHL BL Lac population is more heterogeneous and includes blazars with a broader range of $\nu^{s}_{peak}$. The BL Lacs in 3FHL include 153 LSPs (20\%), 198 ISPs (27\%), 324 HSPs (43\%) and 75 sources with unknown $\nu^{s}_{peak}$ (10\%). These fractions are intermediate between those for blazars found in 3FGL and 2FHL.

% For the 3FGL sources the peak frequencies were adopted from 3LAC \citep{3LAC}. LSP, ISP, and HSP blazars are those with $\log_{10}(\nu^{s}_{peak})<14$, $14<\log_{10}(\nu^{s}_{peak})<15$, $\log_{10}(\nu^{s}_{peak})>15$, respectively.

The 3FHL contains 131 new extragalactic sources. These are typically fainter than the average 3FHL source, and have spectral index similar to the average ($\sim 2.2$). There are 1078 3FHL extragalactic sources detected in the 3FGL, 16 in the 2FHL (and not 3FGL). No 1FHL extragalactic source is missing in the 3FHL catalog. Among all the 3FHL extragalactic sources, 72 have already been detected by IACTs.

\begin{figure}[!ht]
    \centering
    \includegraphics[width=\columnwidth]{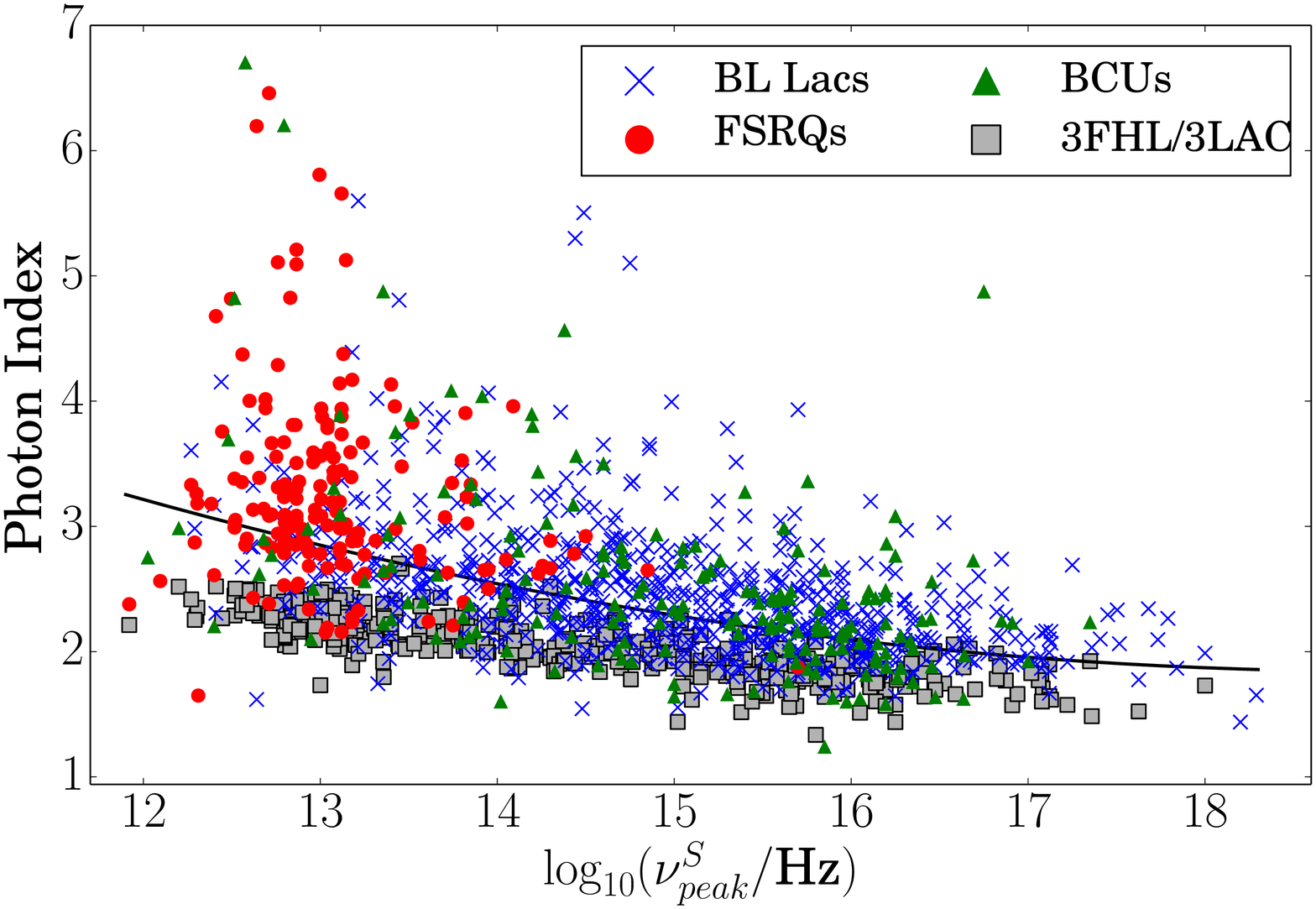}
    \caption{Photon spectral index versus position of the synchrotron peak for the 3FHL blazars (BL Lacs blue crosses, FSRQs red circles, and BCUs green triangles). The best-fitting second-order polynomial to the 3FHL data is shown with a black line ($\chi^{2}/dof=1274/1041$, whereas $\chi^{2}/dof=2700/1043$ for a constant line). The 3LAC data \citep[gray squares,][]{3LAC} of the blazars that are also found in the 3FHL are shown for comparison.
    \label{fig:index_vs_peak}}
\end{figure}

The spectral index is plotted versus the frequency of the synchrotron-peak maximum in Figure~\ref{fig:index_vs_peak}. The trend of a strong hardening of the energy spectra with increasing peak frequency already seen above 100~MeV in the \lat AGN catalogs \citep[\eg][]{3LAC} is even more pronounced above 10~GeV. This enhanced effect relative to 3LAC is due to the larger EBL attenuation suffered by high-redshift sources (most of them being LSPs) in comparison with the lower-redshift ones (preferentially HSPs; see more details on EBL attenuation below). In Figure~\ref{fig:index_vs_peak}, we note one FSRQ that has a hard spectrum ($\Gamma=1.65\pm 0.36$) and low luminosity. This blazar (3FHL J0845.8$-$5551), which should be studied further in a future work, is associated with PMN~J0845$-$5555.

Redshifts have been measured for 548 of the sources (45\% of the extragalactic sample). The median of the redshift distribution is $z\sim 0.4$ but the distribution extends to $z\sim 2.5$. As is well known, BL Lacs typically have lower redshifts (median $z\sim 0.3$) than FSRQs \citep[median $z\sim 1$, ][]{3LAC}. BCUs generally have low redshifts (median $z\sim 0.1$). There are 109 blazars at $z>1$ (82 FSRQs, 31 BL Lacs, and 3 BCUs) and 16 at $z>2$ (11 FSRQs, 4 BL Lacs, and 1 BCUs). We note that only 7 2FHL blazars have redshifts $z>1$.

\begin{figure}[!ht]
    \centering
    \includegraphics[width=8.2cm]{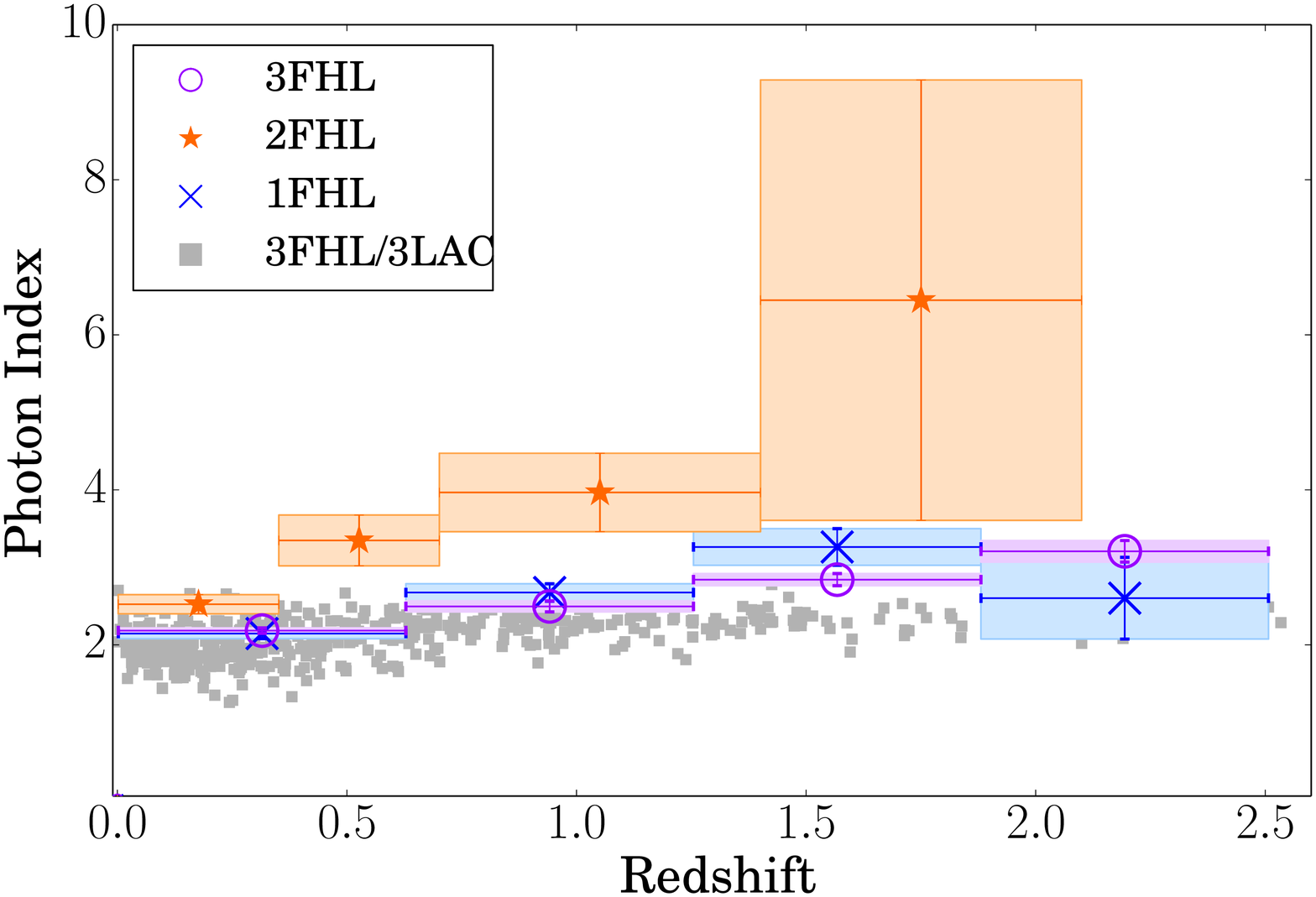} 
    \includegraphics[width=8.2cm]{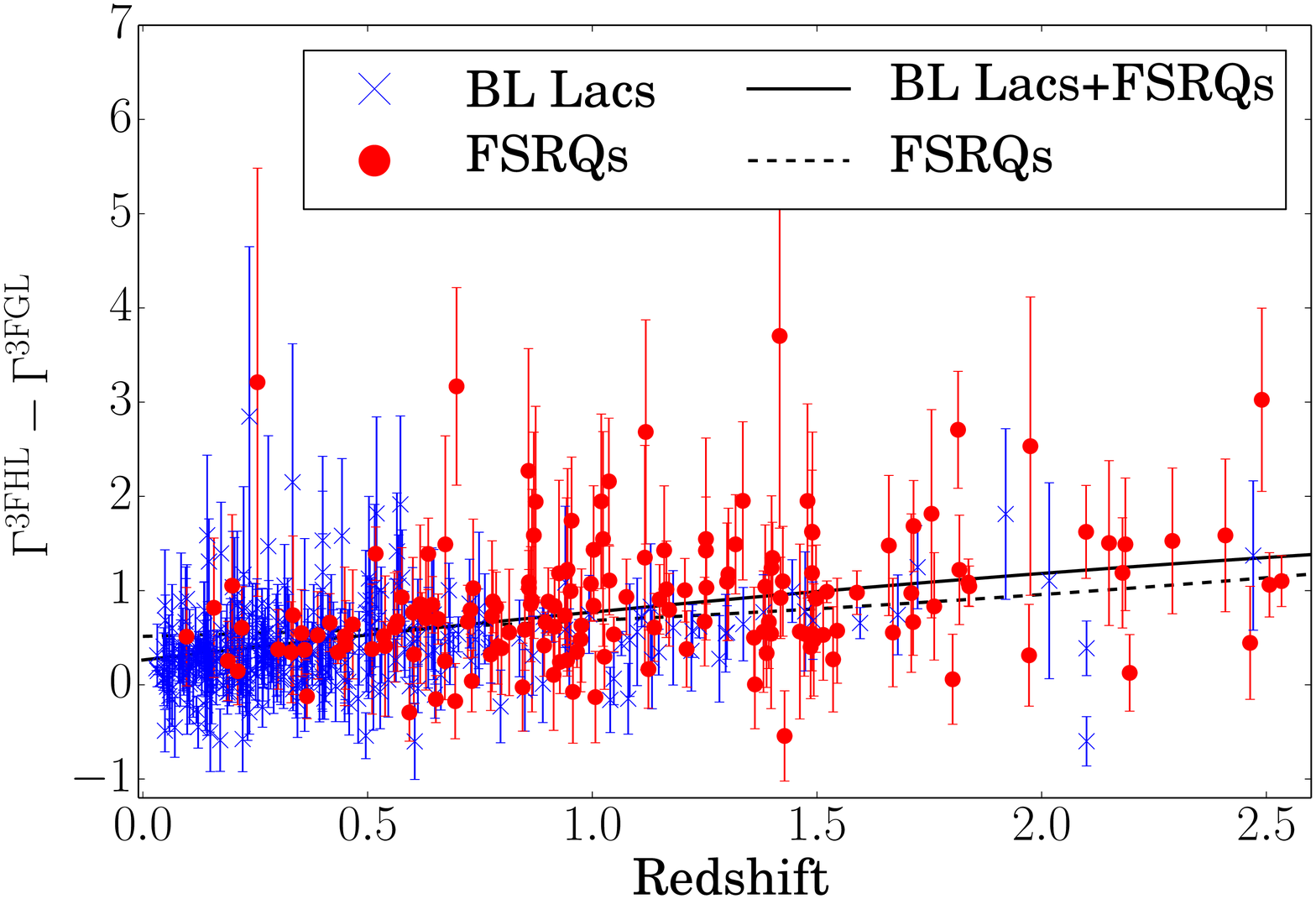} 
    \caption{{\it (Left panel)} Observed spectral index vs. redshift of the 3LAC sources (energy range, 0.1--100 GeV, gray circles), the median spectral index in some redshift bins of the 1FHL sources (10--500~GeV, blue crosses), 2FHL sources (50--2000~GeV, orange stars), and 3FHL sources (10--2000~GeV, magenta circles). The uncertainties are calculated as the 68\% containment around the median. The spectral index is seen to depend on the redshift at the energies where the EBL attenuation is significant. {\it (Right panel)} The difference between the 3FHL and 3FGL spectral index ($\Delta\Gamma=\Gamma^{{\rm 3FHL}}-\Gamma^{{\rm 3FGL}}$) over redshift for the BL Lac (blue crosses) and FSRQ (red circles) populations. The $\Delta\Gamma$ for both population types evolves similarly with redshift.
    \label{fig:index_vs_z}}
\end{figure}

Photons with energies greater than about 30~GeV suffer from attenuation over cosmological distances as a consequence of the pair production interactions with extragalactic background light (EBL) photons \citep[\eg][]{franceschini08,helgason12,scully14}. This interaction results in a cosmic optical depth, $\tau$, to $\gamma$ rays that may be quantified by the cosmic $\gamma$-ray horizon \citep[CGRH, defined as the energy at which $\tau=1$ as a function of redshift, ][]{lat_ebl10,dominguez13a} and carries cosmological information \citep[][]{dominguez13b,biteau15}\footnote{Note that the CGRH is different from the cosmological particle horizon, the distance beyond which information cannot reach the observer.}. Figure~\ref{fig:index_vs_z} shows a clear softening of the spectral index above 10~GeV with increasing redshift, which is likely due to EBL attenuation. This softening was already evident among 1FHL sources. As Figure~\ref{fig:nupeak} illustrated, different mixes of AGN populations dominate each LAT catalog. This fact could introduce some bias in the inferred index evolution; however, we also show in Figure~\ref{fig:index_vs_z} that the difference between the 3FHL and 3FGL spectral index (\ie $\Delta\Gamma=\Gamma^{{\rm 3FHL}}-\Gamma^{{\rm 3FGL}}$) evolves similarly with redshift for the BL Lac as well as FSRQ populations \citep[see also ][]{dominguez15}.

\begin{figure}[!ht]
    \centering
    \includegraphics[width=\columnwidth]{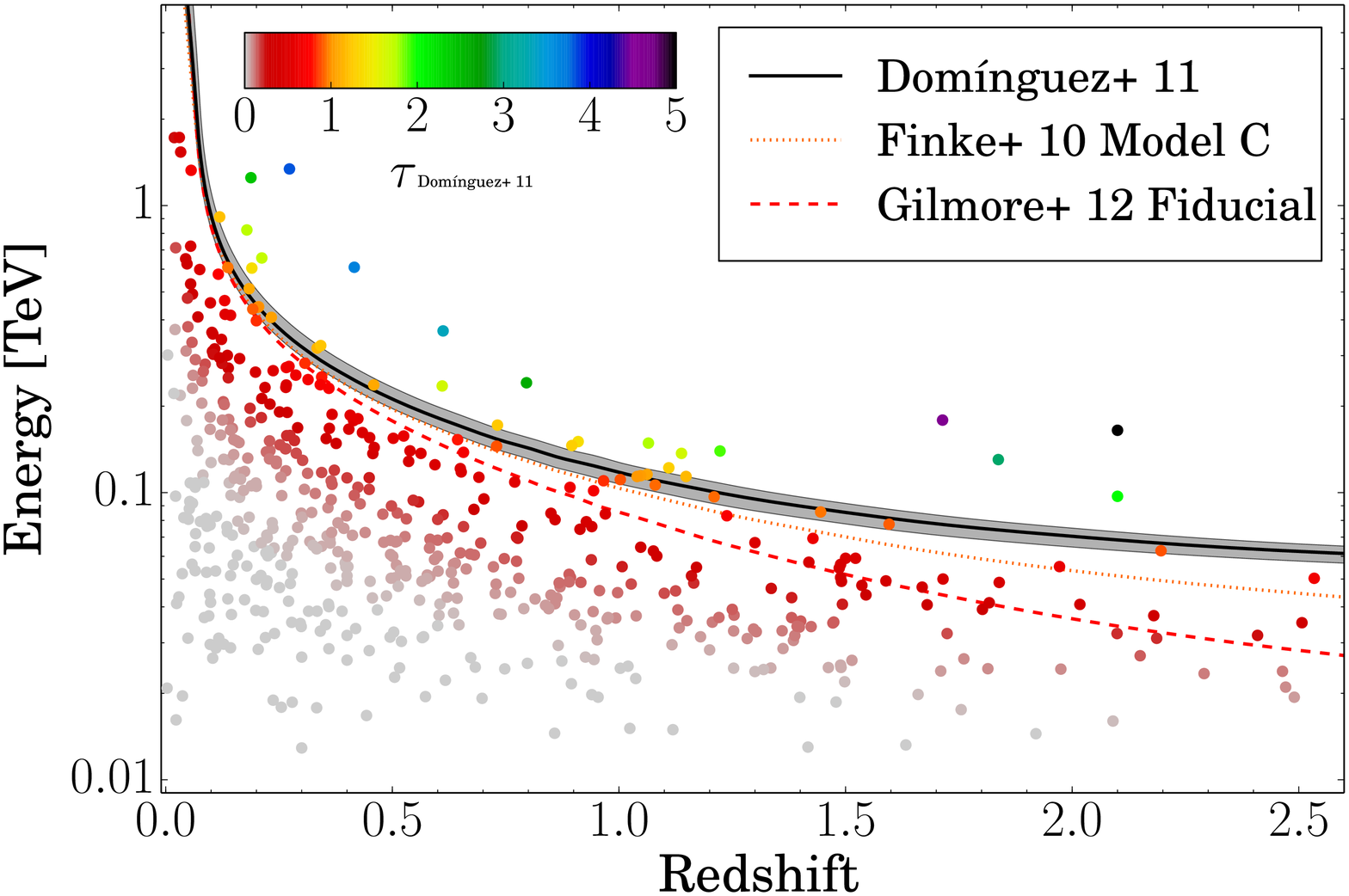} 
    \caption{The highest-energy photons versus redshift for 3FHL associated blazars, color coded by the optical depth, $\tau$, calculated from the model presented by \citet{dominguez11a}. The cosmic $\gamma$-ray horizon (energy for which $\tau=1$ as a function of redshift) from the \citet{dominguez11a}, \citet{finke10}, \citet{gilmore12} EBL models are shown with solid-black line, dotted-orange line, and dashed-red line, respectively. Several of the highest-energy LAT photons from these distant blazars are in the region around and beyond $\tau=1$.
    \label{fig:cgrh}}
\end{figure}

Photons from sources that suffer strong attenuation ($\tau>1$), such as those from 3FHL J0543.9$-$5532 (1RXS~J054357.3$-$553206, $z=0.273$, HEP$_{Energy}=1341$~GeV, $N_{pred}=96$~photons above 10~GeV, HEP$_{Prob}=0.95$), 3FHL J0808.2$-$0751 (PKS 0805$-$07, $z=1.837$, HEP$_{Energy}=130$~GeV, $N_{pred}=80$, HEP$_{Prob}=0.99$), and 3FHL J1016.0+0512 (TXS 1013+054, $z=1.714$, HEP$_{Energy}=179$~GeV, $N_{pred}=31$, HEP$_{Prob}=0.86$), can be powerful probes of the EBL. These photons permit testing EBL models and evaluating potential changes in the optical depth due to other more exotic scenarios \citep[\eg][]{deangelis07,essey10,sanchez-conde09,dominguez11b,horns12}. Figure~\ref{fig:cgrh} shows the HEPs from each source versus redshift together with estimates of the CGRH from EBL models \citep{finke10,dominguez11a,gilmore12}. This figure illustrates that the {\it Fermi}-LAT data explore the region around and beyond the horizon ($\tau=1$).

%%%%%%%%%%%%%%%%%%%%%%%%%%%%%%%%%%%%%%%%%%%%%%%%%%%%%%%%%%%%%%%%
%
%         Variability
%
%%%%%%%%%%%%%%%%%%%%%%%%%%%%%%%%%%%%%%%%%%%%%%%%%%%%%%%%%%%%%%%%
\subsection{Flux Variability}
\label{sec:var}
In this section, we present results on source flux variability following a similar methodology as for previous hard-source LAT catalogs. Our analysis is based on a Bayesian Block algorithm that detects and characterizes variability in a time series. In particular, this analysis can be applied to low-count data \citep{scargle98,scargle13}. The algorithm divides events from a source into blocks (\ie segments of constant rate) described by a flux and duration. The transition between blocks is determined by optimizing a fitness function \citep[using the algorithm of ][]{jackson05}, which in our case is the logarithm of the likelihood for each individual block under the hypothesis of a constant local flux \citep{scargle13}. We select a threshold of 1\% false positive for all sources. Therefore for a catalog of 1556 sources, we expect about 16 false detections. As was done in the 1FHL and 2FHL, we extract the events using an RoI of $0\fdg5$ radius centered on the maximum likelihood source coordinates. For 3FHL sources separated by less than $1^{\circ}$, the radius of the RoI was decreased to the greater of (half the angular separation) or $0\fdg25$. In addition to the Bayesian Block analysis (which accounts for exposure time), we also performed an aperture photometry analysis for each source using 50 equal time bins spanning the whole 3FHL time interval. No background subtraction was done for the aperture photometry and Bayesian Block analyses. Thirty-one pairs of sources are closer than $0\fdg5$, and 8 of those are outside the Galactic plane ($|b|\geq 10^{\circ}$). Of those 8 pairs, 2 pairs show evidence for variability. The variability of the source 3FHL~J1443.5+2515 (NVSS~J144334+251559) cannot be determined independently of the variable source 3FHL~J1443.9+2502 (PKS~1441+25) and is flagged accordingly in the catalog. In the other case 3FHL J1848.5+3217/3FHL J1848.5+3243, only the first object is variable (3 blocks) and it is associated with 3FGL~J1848.4+3216, which was identified with the blazar B2 1846+32A in the 3FGL via correlated multi-wavelength variability. The time of the flare found in the 3FHL Bayesian Blocks analysis matches that of the flare found in the 3FGL analysis (MJD~55500). The second source in the pair is not detected in the 3FGL.

There are 163 sources that are variable (characterized by 2 or more blocks) at more than 99\%~C.L. in the 3FHL catalog. This $\sim$10\% fraction is compatible with the fraction of variable sources found in 1FHL. We also note that 338 3FHL sources were found to be variable in 3FGL. The mean number of detected photons for the subsample of variable sources is 107 whereas for the full catalog it is 48. Naturally, the lower number of photons at higher energies decreases the analysis sensitivity to variability. In the 7-year interval that we analyze, 82 sources have 2 blocks, 50 have 3 blocks, 15 have 4 blocks and 16 have 5 or more blocks. These highly variable sources with 5 or more blocks (including 7 BL Lacs and 9 FSRQs) are associated with B0218+357, PKS~0426$-$380, PKS~0454$-$234, PMN~J0531$-$4827, PKS~0537$-$441, S5~0716+71, Mkn~421, 4C~+21.35, 3C~279, PKS~B1424$-$418, PKS~1510$-$08, TXS~1530$-$131, RX~J1754.1+3212, S4~1849+67, BL~Lacertae and 3C~454.3.

All but two of the variable sources are associated with extragalactic sources. Some other examples of well-known sources that are significantly variable include the local blazars Mkn~180 and Mkn~501, and the radio galaxy IC~310. One unassociated source, 3FHL~J0540.2+0654, is variable (not found variable in 3FGL, located at $b=-12\fdg42$ and therefore probably extragalactic).  One Galactic source is also found to be variable, the extended source 3FHL J1514.2$-$5909e classified as spp (see Table~\ref{tab:classes} for definition). In principle we would not expect an extended source to be variable. Closer study will be needed to understand the origin of its apparent variability. Other interesting cases include the variable source 3FHL J2017.3+0603, which might be associated with the FSRQ GB6 B2014+0553 instead of the adopted association with the pulsar PSR~J2017+0603 (not found variable in the 3FGL).

\begin{figure}[!ht]
    \centering
    \includegraphics[width=\columnwidth]{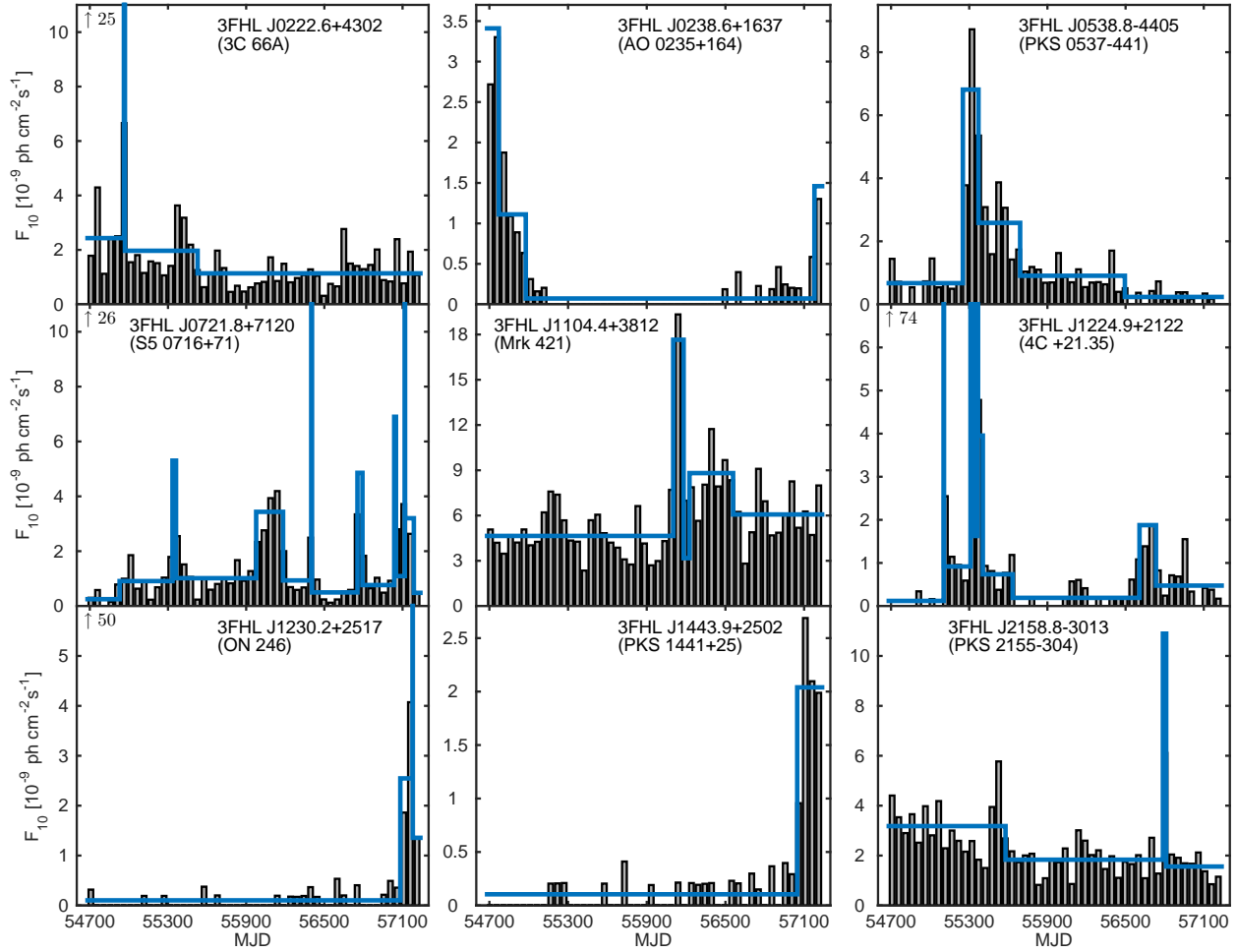} 
    \caption{Light curves for 9 interesting variable sources, which are described in detail in the text. The histograms are from aperture photometry using 50 equal time bins and the solid lines correspond to the Bayesian Block temporal analysis using a 1\% false positive threshold. The $y$-axis is truncated for four sources as the flux estimated by the Bayesian Block analysis was much larger due to the short durations of some flares. An upward-pointing arrow at the top left corner indicates the flux of the brightest flare as estimated by the Bayesian Block analysis. The panels are labelled with the 3FHL names and their corresponding associated sources.
    \label{fig:bb}}
\end{figure}

Figure~\ref{fig:bb} shows the results from the Bayesian Block analysis for nine representative variable sources.
\begin{itemize}
\item 3FHL J0222.6+4302 (3C 66A): We note that one of the flares detected in 1FHL is not picked up in 3FHL. The blazar has been in a quiescent state since the 1FHL time period.
\item 3FHL J0238.6+1637 (AO 0235+164): There is a sign of activity near the end of the 3FHL time window.
\item 3FHL J0538.8$-$4405 (PKS 0537$-$441): The flux  decreased after the 3-year interval analyzed for 1FHL.
\item 3FHL J0721.8+7120 (S5 0716+71): This blazar had two blocks in 1FHL. Now, it is the most variable source in 3FHL with 15 blocks.
\item 3FHL J1104.4+3812 (Mrk 421): This classical TeV source showed a large activity in mid-2012 \citep{hovatta15,balokovic16}, where its flux above 10 GeV increased by about factor of four.
\item 3FHL J1224.9+2122 (4C +21.35): This source is highly variable with 13 blocks. But its variability decreased substantially after the first three years considered in the 1FHL.
\item 3FHL J1230.2+2517 (ON 246): Two flares detected by the LAT, on 2015 January 22, MJD 57044 \citep{atelON246a} and 2015 June 6, MJD 57179 \citep{atelON246b} are clearly seen in the Bayesian Block curves.
\item 3FHL J1443.9+2502 (PKS 1441+25): This source has the second highest redshift among all sources detected so far by IACTs \citep[$z=0.939$, ][]{pks1441veritas,pks1441magic}.
\item 3FHL J2158.8-3013 (PKS 2155-304): This well-known TeV source shows a large flare on 2014 May 17, MJD 56794 \citep{atelPKS2155}.
\end{itemize}

We find that 11\% of sources associated with BL Lacs and 36\% of those associated with FSRQs are flagged as variable. However, the mean number of photons per source for BL Lacs is similar to that of FSRQs. This suggests the interpretation that FSRQs are intrinsically more variable than BL Lacs. The fraction of variable sources is also  found to change with the SED class. It decreases from 23\% (83/360) to 11\% (27/256) and 10\% (44/433) for LSPs, ISPs, and HSPs respectively, confirming the trends seen in 1FHL and 3LAC. If the analysis is restricted to BL Lacs only, the trend is weaker than what was reported in 3LAC. This may be due to BL Lac LSPs having fewer high-energy photons relative to lower energy photons than HSPs while being intrinsically more variable in the overall LAT energy range (as shown in 3LAC).
%%%%%%%%%%%%%%%%%%%%%%%%%%%%%%%%%%%%%%%%%%%%%%%%%%%%%%%%%%%%%%%%
%
%         Summary
%
%%%%%%%%%%%%%%%%%%%%%%%%%%%%%%%%%%%%%%%%%%%%%%%%%%%%%%%%%%%%%%%%
\section{Summary}
\label{sec:summary}

We have analyzed the first 7 years of {\it Fermi}-LAT data using Pass~8 events. Pass~8 improves the photon acceptance and the PSF, reduces the background of misclassified charged particles and extends the useful LAT energy range. Our search for sources above 10~GeV resulted in the 3FHL catalog, which contains 1556 sources characterized up to 2~TeV. This analysis represents a positionally unbiased census of the whole sky at a sensitivity 3 times better than the previous LAT analysis at the same energies (the 1FHL catalog). This improvement in sensitivity results in the detection of 3 times more sources. %\citep[We note that the number of sources in the 3FHL catalog is greater than the number of $\gamma$-rays detected above 10~GeV by the EGRET experiment on the predecessor {\it Compton} Gamma-Ray Observatory mission, \ie 1506 photons, ][]{thompson05}. % To illustrate the improvements brought by the LAT over its predecessor, the Energetic Gamma Ray Experiment Telescope \citep[EGRET,][]{egret99}, we note that the number of sources in the 3FHL is larger than the number of photons detected by EGRET above 10~GeV .

Most of the 3FHL sources ($\gtrsim 79$\%) are associated with extragalactic counterparts. BL Lacs are the most numerous extragalactic population (61\%) followed by blazars of uncertain type (23\%) and FSRQs (14\%).% At $>$10\,GeV blazars are selected regardless of their synchrotron peak position (\ie blazar sub-class).
% We find that at $>$10~GeV and down to our sensitivity limit, there is no a population of blazar-type that clearly dominates over others.

We find 321 sources at $|b|<10^{\circ}$, of which 105 are associated with Galactic-type sources, 133 are extragalactic, and 83 are unassociated (or associated with sources of unknown nature). There are 20 sources with Galactic associations located at high Galactic latitudes ($|b|\geq 10^{\circ}$). About the same number of Galactic sources are associated with PWNe and SNRs as with PSRs. Extragalactic sources generally have smaller photon indices than Galactic ones (median of $\sim 2$ versus $\sim 3$). The unassociated sources tend to have similar indices as blazars, SNRs, and PWNe but not PSRs. Sixty percent of the unassociated sources are located at $|b|\geq 10^{\circ}$ and are likely extragalactic.

The 3FHL catalog contains more than 4 times the number of sources detected in the 2FHL (above 50~GeV). We estimate that down to an energy flux of $10^{-12}$~erg s$^{-1}$cm$^{-2}$, there are twice as many sources at 10~GeV than at 50~GeV. This demonstrates quantitatively the importance of lowering as much as possible the energy threshold of future IACTs.

%%%%%%%%%%%%%%%%%%%%%%%%%%%%%%%%%%%%%%%%%%%%%%%%%%%%%%%%%%%%%%%%%%%%%%%%%%
%%%%%%%%%%%%%%%%%%%%%%%%%%%%%%%%%%%%%%%%%%%%%%%%%%%%%%%%%%%%%%%%%%%%%%%%%%
\acknowledgments
The \textit{Fermi} LAT Collaboration acknowledges generous ongoing support
from a number of agencies and institutes that have supported both the
development and the operation of the LAT as well as scientific data analysis.
These include the National Aeronautics and Space Administration and the
Department of Energy in the United States, the Commissariat \`a l'Energie Atomique
and the Centre National de la Recherche Scientifique / Institut National de Physique Nucl\'eaire et de Physique des Particules in France, the Agenzia 
Spaziale Italiana and the Istituto Nazionale di Fisica Nucleare in Italy, 
the Ministry of Education, Culture, Sports, Science and Technology (MEXT), 
High Energy Accelerator Research Organization (KEK) and Japan Aerospace 
Exploration Agency (JAXA) in Japan, and the K.~A.~Wallenberg Foundation, 
the Swedish Research Council and the Swedish National Space Board in Sweden.
Additional support for science analysis during the operations phase 
is gratefully acknowledged from the Istituto Nazionale di Astrofisica in 
Italy and the Centre National d'\'Etudes Spatiales in France.
This research has made use of the NASA/IPAC Extragalactic Database (NED) which is operated by the Jet Propulsion Laboratory, California Institute of Technology, under contract with the National Aeronautics and Space Administration. This research has made use of the SIMBAD database, operated at CDS, Strasbourg, France

{\it Facilities:} \facility{Fermi/LAT}

%%%%%%%%%%%%%%%%%%%%%%%%%%%%%%%%%%%%%%%%%%%%%%%%%% biblio
\bibliographystyle{apj}
\bibliography{biblio.bib}

\appendix

\section{Background parameters}
\label{app:bkgd}

\begin{figure*}[!ht]
\centering
\includegraphics[width=0.49\textwidth]{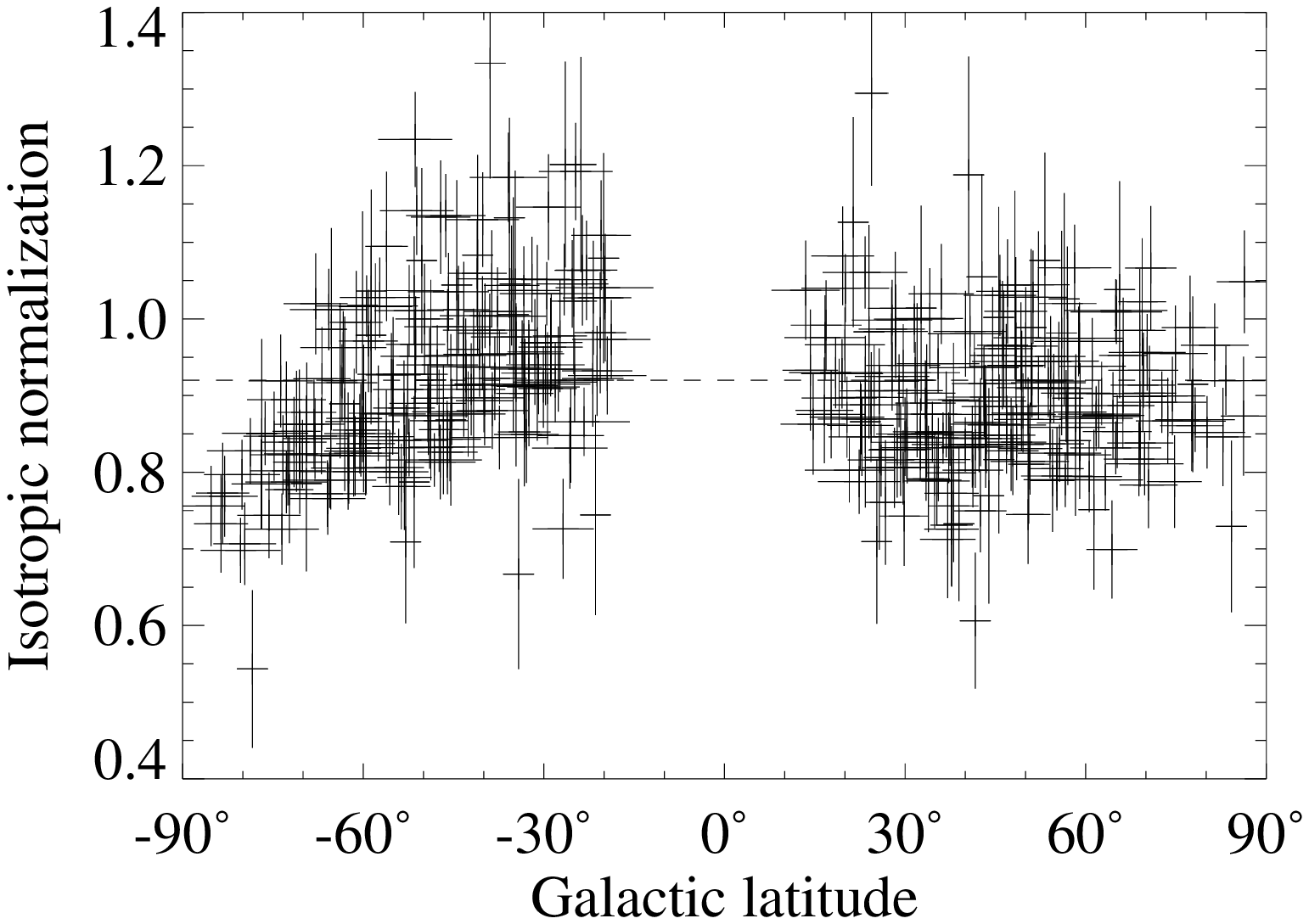}
\includegraphics[width=0.49\textwidth]{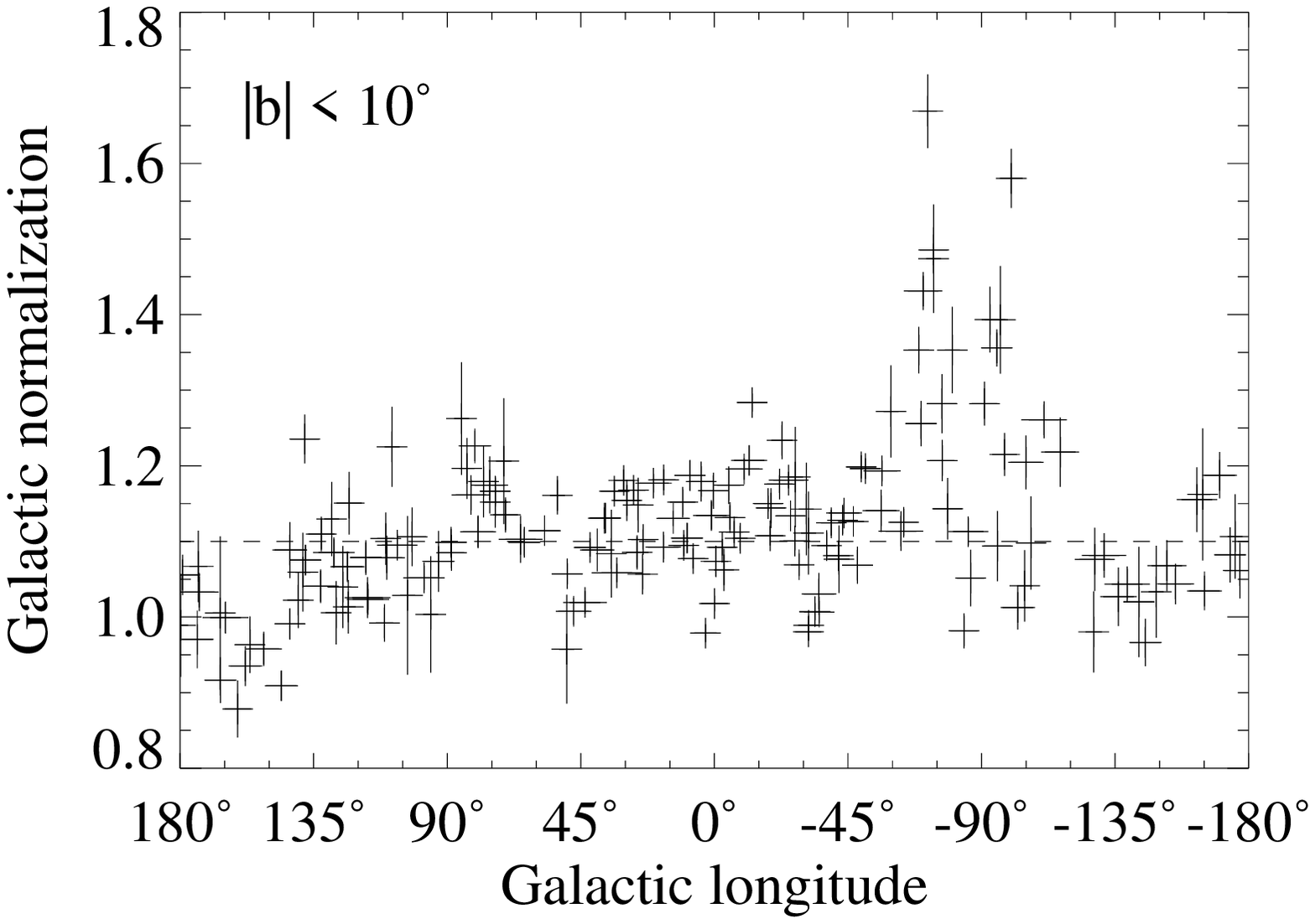}
\caption{Best fit value of the free diffuse parameter. {\it (Left panel)} Isotropic normalization as a function of Galactic latitude. {\it (Right panel)} Galactic normalization (at 20~GeV) within $10\degr$ of the plane as a function of Galactic longitude. The dashed lines (at 0.92 and 1.1 respectively) show what the parameter was fixed to when that component was the minority.}
\label{fig:bkgparams}
\end{figure*} 

In 3FGL we left three degrees of freedom to the diffuse emission in each RoI. Those were the normalizations of the isotropic and Galactic components, and a spectral index $\Gamma$ that can make the Galactic component a little harder or softer after it is multiplied by $(E/E_0)^\Gamma$.
In 3FHL the energy range and statistics are not enough to constrain reliably those three parameters. We first tried with all three free (setting $E_0$ to 20~GeV) but this resulted in large fractional errors (comparable to 1) outside the Galactic plane. In several RoIs one normalization went very close to 0, resulting in too-small error estimates on all parameters (including sources). On the other hand because the PSF is relatively narrow above 10~GeV the effect of background modeling on point sources was much less than over the full LAT energy range, so having a very accurate background model was not as critical.

In order to reach a robust description of the background we adopted the following procedure.
In the Galactic plane ($|b| < 10\degr$) the Galactic parameters were decently measured so we could estimate their averages as norm = 1.1 and $\Gamma$ = 0.03 (the model was a little too faint and too soft). Then we fixed those parameters to those averages and fitted only the isotropic norm. This resulted in an average value of 0.92 outside the plane (model a little too bright). We checked that fixing the isotropic to 0.92 (or to 1) had little effect on the Galactic parameters in the plane. We adopted those values as defaults for the three parameters.

In the final run we left free only one normalization, corresponding to the majority component according to the default parameter values. The other normalization was fixed to its default. The spectral index was always left fixed to 0.03 because there was no evidence that it was significantly variable along the plane.
This resulted in the parameter values illustrated in Fig.~\ref{fig:bkgparams}, which are indeed consistent with the default values on average, and have errors small enough to remain far from 0. 
Among 741 RoIs, 390 had isotropic normalization free and 351 had Galactic normalization free.

\end{document}